\documentclass[prd,aps,a4paper,twocolumn,eqsecnum,floatfix]{revtex4}  

\newif\ifusesec
\usesectrue  
   
\usepackage{graphicx} 
\usepackage{amsmath,amsfonts,amssymb}
\usepackage{mathtools}
\usepackage{color}
\newcommand{\beq}{\begin{equation}}
\newcommand{\eeq}{\end{equation}}
\newcommand{\bea}{\begin{eqnarray}}
\newcommand{\eea}{\end{eqnarray}}

\begin{document}

\title{Radiated Energy Spectrum, Radiated Angular Distribution and Non-linear Memory from the One-loop Gravitational Bremsstrahlung Waveform}

\author{Donato Bini$^{1}$, Thibault Damour$^{2}$, Stefano De Angelis$^3$,  Andrea Geralico$^{1}$}  
  \affiliation{
$^1$Istituto per le Applicazioni del Calcolo ``M. Picone'' CNR, I-00185 Rome, Italy\\
$^2$Institut des Hautes Etudes Scientifiques, 91440 Bures-sur-Yvette, France\\
$^3$Institut de Physique The\'orique, CEA, CNRS, Universit\`e Paris-Saclay,
F-91191 Gif-sur-Yvette cedex, France
}
 
\date{\today}

\begin{abstract}
The frequency-domain gravitational waveform emitted by the scattering of two non-spinning massive particles
has recently been derived at next-to-leading, \textit{i.e.} one-loop, post-Minkowskian order, $h(\omega, \theta,\phi) \sim G^2 + G^3$. 
Building on this one-loop-accurate frequency-domain gravitational waveform, 
we successively derive the spectral gravitational-wave (GW) radiance, $dE^{\rm gw}/(d\omega d\Omega)$,
the radiated GW energy spectrum, $dE^{\rm gw}/d\omega$, and the radiated GW angular distribution, $dE^{\rm gw}/d\Omega$, up to order $G^4$ included.
We deduce from the radiated angular distribution the multipole expansion of the non-linear memory
up to order $G^5$ included, thereby extending previous results.
We work in the center-of-mass frame, and our results reach the fractional 7.5PN accuracy. For completeness, we  include the tree-level information  (considered in the center-of-mass frame).   
\end{abstract}

\maketitle

\section{Introduction}

The emergence of gravitational-wave (GW) astronomy has created a growing demand for high-precision calculations in gravitational physics. The scientific success of present and future GW observatories relies critically on our ability to derive accurate predictions for the dynamics and radiation emitted by compact binary systems within the framework of General Relativity (GR). Achieving this goal is highly nontrivial, due to the non-linear nature of Einstein's equations. As a consequence, one must rely either on numerical relativity methods or on analytical approximation schemes, such as post-Newtonian (PN, weak field and slow motions), post-Minkowskian  (PM, weak field but relativistic motions), PN-matched Multipolar Post-Minkowskian (MPM), and self-force approaches (SF, extreme mass ratio limit of the involved binary system).
In recent years, a new interplay between scattering amplitudes, effective field theory, and classical GR has led to a wealth of new results, considerably improving our predictive power within the PM expansion. 
In practice, all the above approaches are often used synergistically, either to cross-check one another or to complete partial results. Their application to the computation of GW templates used for detecting and analyzing GW signals need further processing or resummation schemes (such as the Effective One-Body approach).
For entries in the literature on analytical approximation schemes, see, e.g., Refs. \cite{Blanchet:2013haa,Buonanno:2022pgc,Bjerrum-Bohr:2022blt,Kosower:2022yvp,Damour:2025uka}.

The primary observable in GW experiments is the gravitational waveform
$f_{ij}^{\rm TT}(t_r, {\bf n})=\lim_{r\to \infty} r h_{ij}^{\rm TT}$, i.e. the radiative component of the gravitational field, $h_{\mu\nu}=g_{\mu\nu}-\eta_{\mu\nu}$, evaluated in the asymptotic region far from the source ($r\to\infty$). The waveform depends on the retarded observation time $t_r$ and on a null vector $n^\mu =(1, {\bf n})$ specifying the direction of the observer.

Building on pioneering studies based on classical techniques, recent developments have established a connection between gravitational waveforms, scattering amplitudes, and diagrammatic methods. 
The state of the art in the knowledge of the bremsstrahlung waveform emitted during the scattering of 
two nonspinning point masses can be summarized as follows:
\begin{enumerate}
\item a PM description at the one-loop ($ h \sim G^3$) level, 
which is given as a closed function to all orders in velocity \cite{Brandhuber:2023hhy,Herderschee:2023fxh,Georgoudis:2023lgf,Brunello:2025eso}. In practice, its high-PN expansion in the cm frame remains largely unexplored due to the technical difficulties involved in extracting PN information from PM results.
See also the more recent work \cite{Heissenberg:2025fcr} which reached the 5PN level in the 1-loop waveform. Here, we extended the PN accuracy of the 1-loop waveform to the 7.5PN level.  

\item a PN-matched MPM description, reaching the fractional 3.5PN accuracy (currently only for the quadrupolar component) and extending up to the $G^3$ (or two-loop) level, with a realistic prospect of further improving the PM accuracy while remaining at 3.5PN order \cite{Bini:2026dvn}; 
 
Let us note that the PN-matched MPM method has reached the 4.5PN level for quasi-circular orbits, and that the corresponding waveform includes 3-loop-level contributions ($ h \sim G^5$) 
\cite{Marchand:2020fpt,Blanchet:2023bwj,Blanchet:2023sbv}. 
\end{enumerate}

The  (frequency-domain) waveform, $\hat f(\omega, {\bf n}) \equiv \lim_{r\to \infty} r \hat h(r,\omega, {\bf n})$ (where ${\bf n}$ denotes the direction of emission),  with its two polarization components,
\beq
\hat f_{+/\times}(\omega, {\bf n}) \equiv \int_{- \infty}^{+\infty} dt_r e^{i \omega t_r} f_{+/\times}(t_r, {\bf n})\,. 
\eeq
gives access to the spectral GW radiance,
\beq
\label{radiance}
\frac{d E_{\rm gw}(\omega, {\mathbf n})}{d \omega \, d\Omega} = \frac{1}{16\pi G}   \frac{\omega^2 }{2\pi}
[ | \hat f_+(\omega, {\bf n})|^2+ | \hat f_\times(\omega, {\bf n})|^2]\,.
\eeq
In the following, we shall generally fold the spectral radiance on the positive frequency axis, and denote it as
\beq   \label{radiance+}
\frac{d E_{\rm gw}^+(\omega, {\mathbf n})}{d \omega \, d\Omega} = 2 \left[ \frac{d E_{\rm gw}(\omega, {\mathbf n})}{d \omega \, d\Omega} \right]^{\omega>0}\,.
\eeq
The knowledge of the waveform at the one-loop level, $\hat f \sim G^2 + G^3$, then gives access to
the spectral GW radiance at the $G^3+G^4$ level. 
The spectral GW radiance is conveniently expanded in spherical harmonics:
\beq
\frac{d E_{\rm gw}^+(\omega, {\mathbf n})}{d \omega \, d\Omega} = \sum_{l\geq 0} \sum_{l=-m}^m \left[\frac{d E_{\rm gw}^+(\omega)}{d \omega}\right]_{l m} Y_{l m}(\theta,\phi)\,.
\eeq
In this paper, we have computed this decomposition and extracted from it some physically-interesting quantities, as we will explain in below. We provide as ancillary files the multipoles $\left[\frac{d E_{\rm gw}^+(\omega)}{d \omega}\right]_{l m}$ as a series of files \texttt{spectrum\_l\_m.m} in the cm frame.

From the spectral GW radiance one can then deduce several physically relevant observables, such as the
GW energy spectrum (with $d\Omega= \sin \theta d\theta d\phi$),
\beq
\frac{d E_{\rm gw}^+(\omega)}{d \omega}= \int d\Omega \frac{d E_{\rm gw}^+(\omega, {\mathbf n})}{d \omega \, d\Omega} \,,
\eeq
and the  radiated GW angular distribution,
\beq
\frac{d E_{\rm gw}({\bf n})}{d \Omega}= \int_0^{+ \infty} d\omega \frac{d E_{\rm gw}^+(\omega, {\mathbf n})}{d \omega \, d\Omega}\,. 
\eeq
The latter quantity is equivalently encoded in its multipole expansion
\beq \label{Elm}
\frac{d E_{\rm gw}({\mathbf n})}{d\Omega}=\sum_{l\ge 0}\sum_{m=-l}^l E^{\rm gw}_{lm}Y_{lm}(\theta,\phi)\,.
\eeq
Evidently, when integrating on all variables one reaches the total losses of energy and linear momentum
\bea
\label{all_E_rad}
E_{\rm rad}&=&\int d\Omega \frac{d E_{\rm gw}({\mathbf n})}{d\Omega}= \sqrt{4\pi} E^{\rm gw}_{00}\,,
\eea
and
\beq
{\bf P}_{\rm rad}= \int d\Omega \, {\bf n}\frac{d E_{\rm gw}({\mathbf n})}{d\Omega}\,,
\eeq
which is encoded in the three $l=1$ coefficients $E^{\rm gw}_{1m}$, with $m=+1, 0, -1$. 
When setting (as in our
previous works) the $x$ axis along the impact parameter, and the $z$ axis along the angular momentum,
the only non vanishing component of ${\bf P}_{\rm rad}$ at $O(G^3)$ is along the $y$ axis while at $O(G^4)$ both the $x$ and $y$ components are present.  We find
\bea
P_x^{\rm rad}&=&\sqrt{\frac{2\pi}{3}}(E^{\rm gw}_{1\bar 1}-E^{\rm gw}_{11})\,,\nonumber\\
P_y^{\rm rad}&=&-i\sqrt{\frac{2\pi}{3}}(E^{\rm gw}_{1\bar 1}+E^{\rm gw}_{11})\,,\nonumber\\
P_z^{\rm rad}&=&0\,.
\eea
We recall that ${E}_{\rm rad}$ and  ${\bf P}_{\rm rad}$ have been  computed (as exact functions
of $p_\infty \equiv \sqrt{\gamma^2-1}$ at both the
LO ($G^3$)  \cite{Herrmann:2021tct}
and the NLO levels ($G^4$) \cite{Dlapa:2022lmu}; at the 2PN fractional accuracy (using the dimensionless angular momentum $j$ in place of the impact parameter $b$, as in the present case) the corresponding expressions were obtained in Ref. \cite{Bini:2021gat}, see Appendices G and H there. 
The  beginning of the PN expansion of these expression are given by
\bea
P_x^{\rm rad}&=& (m_2-m_1)\nu^2 \left(\frac{GM}{b}\right)^4  \pi  P_x^{{\rm rad}\,G^4}\,,\nonumber\\
P_y^{\rm rad}&=& (m_2-m_1)  \nu^2  \left(\frac{GM}{b}\right)^3 \left[ \pi P_y^{{\rm rad}\,G^3}\right.\nonumber\\
&+&\left. \frac{GM}{b}P_y^{{\rm rad}\,G^4}\right]\,, 
\eea
where the (rescaled) $x$ and $y$ components of ${\mathbf P}_{\rm rad}$ are given by
\bea
P_x^{{\rm rad}\,G^4}&=& -\frac{1491}{400} p_\infty^3+\frac{26757}{5600}p_\infty^5-\frac{4046561}{451584}p_\infty^7\nonumber\\
&+&\frac{628523873}{41395200}p_\infty^9
-\frac{3709460060017}{154983628800} p_\infty^{11}\nonumber\\
&+& O(p_\infty^{13})\,.
\eea

\bea
P_y^{{\rm rad}\,G^3}&=&-\frac{37 p_\infty^2}{30}+\left(\frac{37 \nu }{60}-\frac{839}{1680}\right)p_\infty^4\nonumber\\
&+&\left(-\frac{37 \nu ^2}{80}+\frac{107 \nu }{1120}-\frac{2699}{2016}\right) p_\infty^6\nonumber\\
&+&\left(\frac{37 \nu ^3}{96}+\frac{197 \nu ^2}{4480}+\frac{27581 \nu}{40320}+\frac{1531643}{1182720}\right) p_\infty^8\nonumber\\
&+&\left(-\frac{259 \nu ^4}{768}-\frac{715 \nu ^3}{5376}-\frac{14863 \nu ^2}{26880}-\frac{5902873 \nu}{7096320}\right.\nonumber\\
&-&\left. \frac{70348799}{61501440}\right) p_\infty^{10}\nonumber\\
&+& \left(\frac{777 \nu ^5}{2560}+\frac{411 \nu ^4}{2048}+\frac{4775 \nu ^3}{9216}+\frac{7373573 \nu ^2}{9461760}\right.\nonumber\\
&+& \left. \frac{153451423 \nu
   }{184504320}+\frac{30685679}{30750720}\right) p_\infty^{12}\nonumber\\
&+&\left(-\frac{2849 \nu ^6}{10240}-\frac{5253 \nu ^5}{20480}-\frac{19339 \nu ^4}{36864}-\frac{3025135 \nu
   ^3}{3784704}\right.\nonumber\\
&-&\left. \frac{424883111 \nu ^2}{492011520}-\frac{1161339407 \nu }{1476034560}-\frac{6258065657}{7169310720}\right) p_\infty^{14}\nonumber\\
&+& O(p_\infty^{16})\,.
\eea

\bea
P_y^{{\rm rad}\,G^4}&=&-\frac{64}{3}+\left(\frac{32 \nu }{3}-\frac{1664}{175}\right) p_\infty^2 -\frac{128}{3} p_\infty^3\nonumber\\
&+& +\left(-8 \nu ^2+\frac{1096 \nu }{525}-\frac{227776}{33075}\right) p_\infty^4\nonumber\\
&+& +\left(\frac{64 \nu }{3}-\frac{192}{175}\right) p_\infty^5\nonumber\\
&+& +\left(\frac{20 \nu ^3}{3}+\frac{76 \nu ^2}{175}+\frac{118676 \nu }{33075}+\frac{1218176}{72765}\right)p_\infty^6\nonumber\\
&+& \left(-16 \nu ^2-\frac{2512 \nu }{525}-\frac{8528}{189}\right) p_\infty^7\nonumber\\
&+& \left(-\frac{35 \nu^4}{6}-\frac{71 \nu ^3}{35}-\frac{72757 \nu ^2}{22050}-\frac{2297069 \nu }{242550}\right.\nonumber\\
&+&\left. \frac{213323008}{17342325}\right) p_\infty^8\nonumber\\
&+& \left(\frac{40 \nu ^3}{3}+\frac{1328 \nu^2}{175}+\frac{16936 \nu }{675}+\frac{18341432}{363825}\right) p_\infty^9\nonumber\\
&+&\left(\frac{21 \nu ^5}{4}+\frac{97 \nu ^4}{30}+\frac{97201\nu ^3}{26460}+\frac{3980209 \nu ^2}{485100}\right.\nonumber\\
&-&\left. \frac{710768161 \nu }{208107900}-\frac{4728529408}{135270135}\right) p_\infty^{10}\nonumber\\
&+&\left(-\frac{35 \nu ^4}{3}-\frac{338 \nu ^3}{35}-\frac{34201 \nu ^2}{1575}-\frac{1311577 \nu }{40425}\right.\nonumber\\
&-&\left. \frac{234079661}{4729725}\right) p_\infty^{11}\nonumber\\
&+&\left(-\frac{77 \nu ^6}{16}-\frac{1689 \nu^5}{400}-\frac{13267 \nu ^4}{3024}-\frac{336421 \nu ^3}{41580}\right.\nonumber\\
&+&\frac{12417773 \nu ^2}{92492400}+\frac{27089352781 \nu }{1545944400}\nonumber\\
&+&\left.  \frac{2071994415104}{34493884425}\right)p_\infty^{12}\nonumber\\
&+& O(p_\infty^{13})\,.
\eea
We have used these results as checks of our computations.

The main aim of the present paper is to explicitly compute both the multipolar components of
the radiated GW angular distribution,
$E^{\rm gw}_{lm}$ (especially for $l\geq 2$), and the GW energy spectrum $\frac{d E_{\rm gw}^+(\omega)}{d \omega}$. 

It was shown in \cite{BD1989,Blanchet:1990twn,Blanchet:1992br} that the 
decomposition, Eq. \eqref{Elm}, of the angular energy spectrum into {\it scalar} spherical harmonics
gives directly access to the decomposition in (spin-2) {\it tensorial} spherical harmonics of the
{\it nonlinear} memory contribution to the waveform. Namely the $l$th multipolar piece of the
nonlinear memory is given by (here $l \geq 2$)
\beq
\label{hnonlin}
[f_+ - i f_\times] ^{\rm nonlin}_l=16\pi G N_l \sum_{m} E^{\rm gw}_{lm} \,\, {}_{\bar 2}Y_{lm}(\theta,\phi)\,,
\eeq
where the coefficients $E^{\rm gw}_{lm}$ are those of Eq. \eqref{Elm} and where the numerical coefficient
$N_l$ is given by
\beq
N_l=\sqrt{\frac{(l-2)!}{(l+2)!}}\,.
\eeq
For example, $N_2=\frac{1}{2\sqrt{6}}$, $N_3=\frac{1}{2\sqrt{30}}$, $N_4=\frac{1}{6\sqrt{10}}$, etc.

Here the word ``memory" refers to the total shift undergone by the asymptotic gravitational waveform
$f_{ij}^{TT}(t_r, {\bf n})$ as the retarded time $t_r$ evolves from $- \infty$ to $+ \infty$.
We recall that the memory
\beq
[f_{ij}^{TT}] \equiv f_{ij}^{TT}(+ \infty,{\mathbf n}) -  f_{ij}^{TT}(-  \infty,{\mathbf n})
\eeq
is the sum of two contributions: a ``linear memory" term and a ``non-linear memory" one.
The linear memory involves the difference between the
Coulombic metric generated by the outgoing massive particles, and that generated by the ingoing ones
\cite{Gibbons:1971wsk,Zeldovich:1974gvh,Braginsky:1985vlg}. On the other hand, the non-linear memory (usually considered in absence of ingoing radiation)
is associated with the  waveform generated by the outgoing energy-momentum flux of gravitational waves
\cite{BD1989,Blanchet:1990twn,Christodoulou:1991cr,Wiseman:1991ss,Thorne:1992sdb,Blanchet:1992br,Favata:2008yd}. 
It is a direct signature of the nonlinearity of GR, i.e. of the fact that
gravitational waves themselves carry energy and momentum, and therefore gravitate. 
During a binary merger, the nonlinear memory  is approximately proportional to the total emitted gravitational-wave energy and therefore represents an independent observational channel in gravitational-wave astronomy
(which will, however, be challenging to separately measure \cite{Zosso:2026czc}).

The non-linear memory effect is also connected with the infrared structure of gravity, namely
with soft graviton theorems, asymptotic symmetries, and conservation laws at null infinity.
These ideas are part of the so-called \lq\lq infrared triangle" of gravity \cite{Strominger:2013jfa,Strominger:2017zoo}.
In particular, the memory effect is related to the BMS (Bondi-Metzner-Sachs) symmetry group, which describes the asymptotic symmetries of asymptotically flat spacetimes.

Recently, Ref. \cite{Georgoudis:2025vkk} computed the multipole decomposition of the  non-linear memory at leading order in exact PM form in the rest frame of one of the two bodies, while higher-order PN/PM results can be found in Ref. \cite{Bini:2026dvn}. In the present work we shall provide (via our computation of the $E^{\rm gw}_{lm} $
coefficients, when using Eq. \eqref{hnonlin}) 
the non-linear memory at the 1-loop level and at 7.5PN accuracy. To our knowledge, such a result
is new. Reaching such a high PN order is not merely an exercise in technical virtuosity: it allows us to explore terms originating from radiation reaction  ten velocity orders  beyond its first appearance at 2.5PN, which are combined with  high-velocity corrections  to tails associated with nonlocal contributions. 

The structure revealed by our high-PN expansion highlights the presence of patterns that are expected to appear in related computations, such as that of the radial action.
 
Throughout this work we use the mostly positive metric signature ($-+++$) and set $c=1$, unless stated otherwise. We consider the scattering of two masses $m_1$ and $m_2$ (with relative Lorentz factor 
$\gamma = - u_1 \cdot u_2$) moving in the $x-y$ plane,
with impact parameter $b=b_1-b_2>0$ directed along the $x$ axis. We denote $M = m_1+m_2$,
$\nu = m_1 m_2/M^2$, $X_1= m_1/M$,  $X_2= m_2/M$, $X_{21} \equiv X_2-X_1 \equiv \frac{m_2-m_1}{M}$. We denote $p_\infty = \sqrt{\gamma^2-1}$, and we assume that $m_2 >m_1$ so that 
$X_{21} = \sqrt{1-4\nu} >0$.

\section{ Radiated GW angular distribution and Non-linear Memory}

In the following, we find convenient to work with the complex version of the waveform defined by contracting $f_{ij}^{TT}$ with the null polarization vector 
\bea
\bar m&=& \frac1{\sqrt{2}} \left( \frac{\partial{\bf n}}{\partial \theta} -\frac{i}{\sin \theta}\frac{\partial{\bf n}}{\partial \phi} \right)\nonumber\\ 
&=&  \frac1{\sqrt{2}} ( e_\theta - i e_\phi)\,,
\eea
which is orthogonal to the observation direction 
\beq
{\mathbf n}\equiv (n^1,n^2,n^3)=(\sin \theta \cos\phi, \sin \theta \sin \phi, \cos\theta)\,.
\eeq
It is also sometimes convenient to rescale the complex waveform by a factor $(4 G)^{-1}$, i.e.
\bea
\bar m^\mu  \bar m^\nu f_{\mu\nu}= f_+ - i f_\times \equiv 4G W(t_r,{\mathbf n})\,.
\eea
As in our previous works \cite{Bini:2026dvn,Bini:2024ijq,Bini:2024rsy,Bini:2023fiz} we work in the cm frame and use as $z$ axis the direction of the
orbital angular momentum, and as $x$ axis the direction of the eikonal  
impact parameter,
oriented from particle 2 to particle 1. [For more details, see Section II of \cite{Bini:2024rsy}; note that the spectral radiance considered below is insensitive to supertranslations, and in particular insensitive to the cm position difference between the MPM and EFT frames recently discussed in \cite{Bini:2026dvn}.]
The unit vector ${\mathbf n}$ characterizes the spatial propagation direction of the gravitational wave and is associated, in the cm frame, with the null vector
\beq
k=\omega\,  (1,{\mathbf n})\,.
\eeq
The spectral radiance of gravitational radiation at infinity (folded on the positive frequency axis),
see Eqs. \eqref{radiance}, \eqref{radiance+}, can be rewritten as follows in
terms of the waveform in frequency space:
\begin{equation}
\label{dE_domega_dOmega}
\frac{d E^+_{\rm gw}}{d \omega d\Omega}(\omega,{\mathbf n}) = \frac{\omega^2}{16 \pi^2 G}\,  \left[  
|\hat f_+ - i \hat f_\times|^2_{\omega>0}+  |\hat f_+ +i \hat f_\times|^2_{\omega>0} \right] ,
\end{equation}
where the righthand side involves a sum over the two helicity states of the gravitational wave.
The latter sum replaces the factor 2 in   Eq. \eqref{radiance+}. 

From $ \hat f$ at $G^2+G^3$  (i.e. $W$ at $G^1+G^2$), we can compute the gravitational-wave energy spectrum at $G^3 +G^4$ (i.e. the nonlinear memory at $G^4 +G^5$).
 We work in the post-Newtonian expansion, up to the {\it fractional} 7.5PN accuracy
 (i.e., a factor $p_\infty^{15}$ beyond the LO result), and in the center-of-mass frame. This will be the central quantity from which we will obtain the non-linear memory.  In the following, we will introduce the dimensionless rescaled frequency
\begin{equation}
    u = \frac{\omega b}{p_\infty} \,,
\end{equation}
which should not be confused with the standard notation for the retarded time (which we instead denote $t_r$).

Using this notation, the linear memory reads
\beq
\label{weinb}
[W]^{\rm lin}({\mathbf n})= \left[\sum_{a=1}^2  \frac{(p_a\cdot \bar m)^2}{E_a -{\mathbf p}_a \cdot  {\mathbf n}}\right]_{-\infty}^{+\infty}\,,
\eeq
while the ``non-linear memory'' reads (when assuming that the initial state does not contain incoming gravitons)
\bea
[W]^{\rm nonlin}({\mathbf n})&=&\int_{-\infty}^\infty dt_r \int d\Omega' \frac{dE_{\rm gw}(t_r,{\mathbf n}')}{dt_r \, d\Omega' }\,\, \frac{({\mathbf n}'\cdot \bar m)^2}{1-{\mathbf n}'\cdot {\mathbf n}}\,.\nonumber\\
\eea
 
We decompose the total memory (i.e., the sum of the linear and nonlinear contributions) $[W]^{\rm tot}({\mathbf n})$, which is a spin-weight $-2$ field on the sphere, in spin-weighted spherical harmonics, using the  notation ${}_{-2}Y_{lm}(\theta,\phi) \equiv {}_{\bar 2}Y_{lm}(\theta,\phi)$:
\beq
[W]^{\rm tot}({\mathbf n})=\sum_{l\ge 2} [W]^{\rm tot}_l\,,
\eeq
where 
\beq
\label{W_tot_fin}
[W]^{\rm tot}_l=\sum_{m=-l}^l \left(\int d\Omega' W^{\rm tot}({\mathbf n}')\,\,{}_{\bar 2}Y_{lm}^*(\theta',\phi')\right) {}_{\bar 2}Y_{lm}(\theta,\phi)\,.
\eeq

Let us rewrite Eq. \eqref{dE_domega_dOmega} as
\bea
\label{dE_domega_dOmega}
\frac{d E^+_{\rm gw}}{d \omega d\Omega}(\omega,{\mathbf n}) &=& \frac{\omega^2}{16 \pi^2 G}\,  
\left[|\bar m^\mu  \bar m^\nu \hat f_{\mu\nu}(\omega, {\bf n})|^2_{\omega>0} \right.\nonumber\\
&+&\left. | m^\mu  m^\nu \hat f_{\mu\nu}(\omega, {\bf n})|^2_{\omega>0} \right]\,.
\eea
The first term on the righthand side involves  the spin weight $-2$ polarization, while the second one  
involves the spin weight $+2$.
One passes from the first to the second polarization by changing $\bar m^\nu$ into $ m^\nu$.
Note that the second polarization is {\it not} equal to the complex conjugate of the
first because $\hat f_{\mu\nu}(\omega)$ is complex, rather than real. In fact, as 
$\hat f_{\mu\nu}(\omega)$ is the Fourier transform of the real function $ f_{\mu\nu}(t_r)$
we have
\beq
(\hat f_{\mu\nu}(\omega))^*=\hat f_{\mu\nu}(-\omega) \neq \hat f_{\mu\nu}(\omega) \,.
\eeq
Therefore, while the first term is expressible in terms of the complex, rescaled waveform 
$\hat W(\omega)$ for $\omega>0$,
one cannot directly express the second term in terms of the complex conjugate of  $\hat W(\omega)$
evaluated for $\omega>0$.

The first term in the above equation,  proportional to
$| \bar m^\mu  \bar m^\nu \hat f_{\mu\nu}(\omega, {\bf n}) |^2$,
is a function of $\omega$, $\theta$, and $\phi$.
Let us now show that the second term (based on the spin-weight $+2$ polarization)  in the latter equation can be obtained by replacing in the first term the angles $\theta$, $\phi$ by suitably transformed values $\theta'$, $\phi'$. [This operation leaves invariant the frequency, and is therefore valid when considering
only positive frequencies.]

Indeed, as was recently pointed out \cite{Heissenberg:2025fcr}, the following change of variables,
\beq
\theta'=\theta+\pi\,,\qquad \phi'=\phi+\pi\,  ,
\eeq
has the effect of transforming the spin-weight $-2$ quantity
$\hat f_{-2}(\omega, \theta, \phi) \equiv  \bar m^\mu  \bar m^\nu \hat f_{\mu\nu}(\omega, {\bf n})$
into its  spin-weight $+2$ counterpart, $\hat f_{+2}(\omega, \theta, \phi) \equiv   m^\mu   m^\nu \hat f_{\mu\nu}(\omega, {\bf n})$, namely
\beq
\hat f_{-2}(\omega, \theta', \phi')= \hat f_{+2}(\omega, \theta, \phi)\,.
\eeq

This property relies both on the transformation properties of $\bar m^\mu=(0, \bar m^i)$
and $n^\mu=(1, n^i)$ under the mapping
\beq
\label{map_theta_phi}
\theta\to \theta'=\theta+\pi\,,\qquad \phi\to \phi'=\phi+\pi\,,
\eeq
and on the special structure of the waveform $\hat f_{-2}(\omega, \theta, \phi) \equiv  \bar m^\mu  \bar m^\nu \hat f_{\mu\nu}(\omega, {\bf n})$.

First, let us consider the effect of the transformation \eqref{map_theta_phi} on $n^i$ and $\bar m^i$.
Using the  abbreviated notation $[\cos\alpha, \sin \alpha]=[c_\alpha, s_\alpha]$, the explicit expressions
of  
$\bar m^i$ as function  
of $\theta$ and $\phi$  
is
\bea
\bar m^i(\theta,\phi)&\equiv & (\bar m^1,\bar m^2, \bar m^3)\nonumber\\
&=&\frac{1}{\sqrt{2}}(c_\theta c_\phi+is_\phi, c_\theta s_\phi-ic_\phi, -s_\theta)\,.
\eea
It is then easily checked that under the mapping  \eqref{map_theta_phi}, we have
\bea
n^1(\theta',\phi')&=& n^1(\theta,\phi)\,,\quad n^2(\theta',\phi') =n^2(\theta,\phi)\,,\nonumber\\
n^3(\theta',\phi')&=& -n^3(\theta,\phi)\,,
\eea
and
\bea
\bar m^1(\theta',\phi')&=& \bar m^1(\theta,\phi)\,,\quad \bar m^2(\theta',\phi') =\bar m^2(\theta,\phi)\,,\nonumber\\
\bar m^3(\theta',\phi')&=& -\bar m^3(\theta,\phi)\,.
\eea
The transformation  \eqref{map_theta_phi} thus effects  a reflection of both vectors in the orbital plane 
$x-y$.

Lorentz invariance, and the physical symmetry under the orbital plane (when considering spinless bodies) implies that the polarisation vectors are always contracted with one of the four-velocity vectors $u_a^\alpha$ or the impact parameter $b^\alpha$, which span the scattering plane. Thus, the component $m^3=m^z$ does not appear in the waveform. One can also show that $n^3=n^z$ does not enter the waveform. This can be
seen either by considering
the  MPM  description of the waveform in terms of Cartesian tensors (which are finally reduced to tensors 
having only indices in the $x-y$  plane), or by considering the effect of the integration on the
momentum transfer $q$ entering the impact-parameter transform (where the integral on $q^z$ has
no associated exponential factor and therefore cannot select among the two possible orientations
of the $z$ axis orthogonal to the orbital plane).

It is computationally convenient to replace the trigonometric functions of $\theta$ and $\phi$ 
entering the waveform by the variables
\beq
y \equiv e^{i\theta}\,, \qquad z \equiv e^{i\phi}\,,
\eeq
in terms of which the waveform becomes a Laurent polynomial in $y$ and $z$.
The effect of the map \eqref{map_theta_phi} on these variables is
\beq
\label{map_yz}
y\to -y\,,\qquad z\to -z\,.
\eeq

\section{Results for the nonlinear memory in the cm frame}

As recalled in the Introduction, the computation of the multipole expansion of the nonlinear memory
is equivalent (modulo a factor $16 \pi G N_l$, see Eq. \eqref{hnonlin}) to the computation of
the multipole expansion of the angular distribution of GW energy $\frac{dE^{\rm gw}(\Omega)}{d\Omega}$.

We computed the multipole coefficients $E_{lm}^{\rm cm}$ of the latter quantity
in the cm-frame at the two- and three-loop level, i.e.
\bea
\frac{dE^{\rm gw}(\Omega)}{d\Omega}\bigg|_{\rm cm}= \sum_{l\ge 0} \sum_{m=-l}^l E_{lm}^{\rm cm} Y_{lm}(\theta,\phi)\,,
\eea
where $E_{lm}^{\rm cm} = O(G^3+G^4)$ is naturally decomposed as
\beq
E_{lm}^{\rm cm}=E_{lm}^{\rm tree}+E_{lm}^{\rm 1loop}\,.
\eeq
Here each contribution is obtained as a power-series expansion in the velocity parameter
$p_\infty \equiv \sqrt{\gamma^2-1}$ (where $\gamma = - u_1 \cdot u_2$ is the relative Lorentz factor).

Factoring out some convenient prefactors, we write
\bea
E_{lm}^{\rm tree}&=& \frac{G^3m_1^2m_2^2}{b^3} \sum_{k=-1}^{N_{\rm tree}}   p_\infty^k  E^{{\rm cm \, tree} }_{lm(k)}\,,\nonumber\\
E_{lm}^{\rm 1loop}&=& \frac{G^4m_1^2m_2^2 M}{b^4} \sum_{k'=-3}^{N_{1\rm loop}} p_\infty^{k'}  E^{{\rm cm \, 1loop} }_{lm( k' )}\,.
\eea
The maximal power of $p_\infty$ can differ between the tree-level contribution, which is fully known and easily PN-expandable, and the 1-loop contribution, which is formally fully known but not easily PN-expandable. 

In what follows, we further rescale both of them, using a different rescaling for each pair $(l,m)$ in order to remove overall factors of $GM$, $\nu$, and $b$, as well as factors coming from the definitions of the spherical harmonics and the extra factor of $2$ arising from the integration over positive and negative frequencies together, which doubles the result when $l+m$ is even and cancels it when $l+m$ is odd.
At 1-loop order we also split the $E_{lm}$ coefficients into their real and imaginary parts, the latter carrying an additional factor of $\pi$.

We have computed $E_{l}^{\rm cm}$ at the fractional 7.5PN accuracy.
This corresponds to the maximum powers $N_{\rm tree}=14$ and $N_{1\rm loop}=12$ in the waveform. 
$E_{lm (k)}^{{\rm cm} }$ vanishes when  $l+m$ is odd and $k+l$ is even. 
Moreover, the complex conjugate of $E_{lm(k)}^{{\rm cm} }$ satisfies 
\beq
\left(E_{lm(k)}^{{\rm cm} } \right)^*=(-)^m E_{l\bar m (k)}^{{\rm cm} }\,,
\eeq
where $\bar m \equiv - m$ (with a slight abuse of notation, $\bar m$ here denotes the negative of the magnetic number $m$ and not, as above, the polarization vector). It is therefore sufficient to display
the coefficients $E_{lm (k)}^{{\rm cm} }$ for $m\geq 0$.
$E_{lm (k)}^{{\rm cm \, tree,1loop} }$ is either real or purely imaginary at tree level, while at 1-loop order it is in general complex, with a factor of $\pi$ accompanying the imaginary part.

For example, for $l=2$ and $m=2,0$, the above rescalings are implemented through the following notation:
\begin{widetext}
\bea
E_{22}^{\rm tree}&=& \frac{ G^3 M^4 \nu^2}{b^3}  \sqrt{\frac{\pi}{30}}{\mathcal E}_{22}^{\rm tree}(p_\infty,\nu)\,,\nonumber\\
E_{22}^{\rm 1loop}&=& \frac{ G^4 M^5 \nu^2}{\pi b^4}\sqrt{\frac{\pi}{30}}[{\mathcal E}_{22}^{{\rm 1loop}\not \pi}(p_\infty,\nu)
+i\pi {\mathcal E}_{22}^{{\rm 1loop}\pi}(p_\infty,\nu)]\,,\nonumber\\
E_{20}^{\rm tree}&=& \frac{ G^3 M^4 \nu^2}{b^3}  \sqrt{\frac{\pi}{5}}{\mathcal E}_{20}^{\rm tree}(p_\infty,\nu)\,,\nonumber\\
E_{20}^{\rm 1loop}&=& \frac{ G^4 M^5 \nu^2}{\pi b^4}\sqrt{\frac{\pi}{5}}[{\mathcal E}_{20}^{{\rm 1loop}\not \pi}(p_\infty,\nu)
+i\pi {\mathcal E}_{20}^{{\rm 1loop}\pi}(p_\infty,\nu)]\,.
\eea
\end{widetext}
Note that ${\mathcal E}_{21}^{\rm tree}$ and ${\mathcal E}_{2m}^{{\rm 1loop}}$ vanish identically.

The coefficients ${\mathcal E}_{2m}^{\rm tree}$, ${\mathcal E}_{2m}^{{\rm 1loop}\not \pi}$, and ${\mathcal E}_{2m}^{{\rm 1loop}\pi}$ for $m=2,0$ are listed in Tables \ref{tab1:Elm} and \ref{tab2:Elm}, respectively. The coefficients with negative $m$ are the complex conjugates of the corresponding positive-$m$ ones and are not displayed.
\begin{table*}
\caption{\label{tab1:Elm}  Quadrupolar $(l=2, m=2)$  GW energy multipolar  coefficients at order $O(G^5)$.}
\begin{ruledtabular}
\begin{tabular}{ll}
${\mathcal E}_{22}^{\rm tree}$ & $\frac{2}{7} p_\infty+\left(\frac{253 \nu }{32}-\frac{2489}{448}\right)p_\infty^3
\left(-\frac{12007 \nu ^2}{2464}+\frac{1777 \nu}{4928}+\frac{94487}{29568}\right) p_\infty^5$\\
&$+\left(\frac{943505 \nu^3}{256256}+\frac{82081 \nu ^2}{512512}+\frac{1191187 \nu}{109824}-\frac{35436295}{6150144}\right) p_\infty^7$\\
&$+\left(-\frac{389995\nu ^4}{128128}-\frac{113385 \nu ^3}{146432}-\frac{6456075 \nu^2}{1025024}-\frac{46907335 \nu}{4100096}+\frac{36494587}{6150144}\right)
   p_\infty^9$\\
&$+\left(\frac{184121491 \nu ^5}{69701632}+\frac{180048763 \nu^4}{139403264}+\frac{1025608223 \nu^3}{209104896}+\frac{1101879563 \nu^2}{139403264}+\frac{3000851585 \nu
   }{278806528}-\frac{667987769}{119488512}\right)p_\infty^{11}$\\
&$+\left(-\frac{3127604621 \nu^6}{1324331008}-\frac{1144232975 \nu^5}{662165504}-\frac{70451964265 \nu^4}{15891972096}-\frac{55583593451 \nu^3}{7945986048}-\frac{43346987825 \nu^2}{5297324032}-\frac{104206696961 \nu}{10594648064}+\frac{2547233861}{496624128}\right)
   p_\infty^{13}$ \\
&$+O(p_\infty^{14})$\\
\hline
${\mathcal E}_{22}^{{\rm 1loop}\not \pi}$ &$ \frac{80}{21 p_\infty}+\left(\frac{3400 \nu }{21}-\frac{76616}{735}\right)p_\infty
+\frac{160 p_\infty^2}{21}
+\left(-\frac{1690 \nu^2}{21}-\frac{4570 \nu }{147}+\frac{1589842}{24255}\right)p_\infty^3+\left(\frac{6800 \nu }{21}-\frac{21568}{105}\right)p_\infty^4$\\
&$
+\left(\frac{1265 \nu ^3}{21}-\frac{964 \nu^2}{105}+\frac{10285369 \nu}{88935}-\frac{8767559}{385385}\right) p_\infty^5+\left(-\frac{3380 \nu ^2}{21}-\frac{668 \nu}{7}+\frac{77216}{495}\right) p_\infty^6$\\
&$
+\left(-\frac{8425\nu ^4}{168}-\frac{493667 \nu ^3}{140140}-\frac{2363360887 \nu^2}{72144072}-\frac{74919066319 \nu}{360720360}+\frac{853121219}{72144072}\right)
   p_\infty^7$\\
&$+\left(\frac{2530 \nu ^3}{21}+\frac{6514 \nu^2}{105}+\frac{170386 \nu }{385}-\frac{15997952}{63063}\right) p_\infty^8$\\
&$+\left(\frac{2105 \nu ^5}{48}+\frac{4948193 \nu^4}{336336}+\frac{429564589 \nu ^3}{16032016}+\frac{69874600043\nu ^2}{1082161080}-\frac{1831502483 \nu}{721440720}+\frac{419019785537}{4088164080}\right)p_\infty^9$\\
&$+\left(-\frac{8425 \nu^4}{84}-\frac{14723 \nu ^3}{210}-\frac{3060461 \nu^2}{13860}-\frac{96859657 \nu}{180180}+\frac{89831296}{315315}\right) p_\infty^{10}$\\
&$+\left(-\frac{2525 \nu ^6}{64}-\frac{2714112613 \nu^5}{114354240}-\frac{234855863059 \nu^4}{8176328160}-\frac{17648848863211 \nu^3}{312744552120}+\frac{103458128669521 \nu^2}{2501956416960}+\frac{551575279139147 \nu}{2501956416960}-\frac{11694079237103761}{47537171922240}\right)p_\infty^{11}$\\
&$+\left(\frac{2105 \nu ^5}{24}+\frac{323 \nu^4}{4}+\frac{722731 \nu ^3}{3960}+\frac{739166837 \nu^2}{2522520}+\frac{202998947 \nu}{360360}-\frac{4658172928}{16081065}\right)p_\infty^{12}+O(p_\infty^{13})$\\
${\mathcal E}_{22}^{{\rm 1loop} \pi}$&$ \left(\frac{26213 \nu }{560}-\frac{31761}{2240}\right) p_\infty^4
+\left(-\frac{280801 \nu ^2}{49280}-\frac{216081403 \nu}{2759680}+\frac{2149031}{107520}\right) p_\infty^6$\\
&$+\left(\frac{541077\nu ^3}{320320}+\frac{29733563 \nu ^2}{3843840}+\frac{30655433923\nu }{215255040}-\frac{592378823}{16558080}\right)p_\infty^8$\\
&$+\left(-\frac{89891 \nu ^4}{128128}-\frac{698889199 \nu^3}{344408064}-\frac{36806332709 \nu^2}{2583060480}-\frac{1251942322261 \nu}{5166120960}+\frac{53233300271}{861020160}\right)p_\infty^{10}$\\
&$+\left(\frac{136753 \nu ^5}{388960}+\frac{4695896981 \nu^4}{5854937088}+\frac{57857658121 \nu^3}{15968010240}+\frac{981483511837 \nu^2}{39032913920}+\frac{1944868799921 \nu}{5018517504}-\frac{2102145664181}{20664483840}\right)p_\infty^{12}+O(p_\infty^{13}) 
$\\
\end{tabular}
\end{ruledtabular}
\end{table*}
\begin{table*}
\caption{\label{tab2:Elm}  Quadrupolar $(l=2, m=0)$ GW energy multipolar  coefficients at order $O(G^5)$.}
\begin{ruledtabular}
\begin{tabular}{ll}
${\mathcal E}_{20}^{\rm tree}$ & $\frac{73}{42} p_\infty+\left(\frac{545 \nu }{168}+\frac{227}{336}\right)p_\infty^3 +\left(-\frac{359 \nu ^2}{192}+\frac{54421 \nu}{59136}-\frac{21463}{88704}\right) p_\infty^5$\\
&$+\left(\frac{523165 \nu^3}{384384}-\frac{77911 \nu ^2}{384384}+\frac{856433 \nu
   }{288288}-\frac{737101}{838656}\right) p_\infty^7$\\
&$+\left(-\frac{563677\nu ^4}{512512}-\frac{45025 \nu ^3}{292864}-\frac{5513327 \nu
   ^2}{3075072}-\frac{79749469 \nu}{24600576}+\frac{6233659}{5677056}\right)p_\infty^9$\\
&$+\left(\frac{12288335 \nu ^5}{13069056}+\frac{40357171 \nu^4}{104552448}+\frac{113761411 \nu ^3}{78414336}+\frac{969612053
   \nu ^2}{418209792}+\frac{321176533 \nu}{104552448}-\frac{2764581359}{2509258752}\right)p_\infty^{11}$\\
&$+\left(-\frac{347878645 \nu^6}{418209792}-\frac{17799952447 \nu^5}{31783944192}-\frac{64594146553 \nu^4}{47675916288}-\frac{1935203593 \nu^3}{916844544}-\frac{51199276595 \nu^2}{21189296128}-\frac{238156185269 \nu }{84757184512}+\frac{198432769855}{190703665152}\right)p_\infty^{13}$\\
&$+O(p_\infty^{14})$\\
\hline
${\mathcal E}_{20}^{{\rm 1loop}\not \pi}$ &$\frac{1552}{63 p_\infty}
+\left(\frac{4136 \nu }{63}+\frac{488}{35}\right)p_\infty
+\frac{3104}{63}p_\infty^2
+\left(-\frac{1874 \nu^2}{63}+\frac{8006 \nu }{2205}-\frac{632402}{24255}\right)p_\infty^3$\\
&$+\left(\frac{8272 \nu }{63}-\frac{320}{21}\right)p_\infty^4
+\left(\frac{1357 \nu ^3}{63}-\frac{572 \nu^2}{63}+\frac{594277 \nu }{72765}+\frac{526859}{21021}\right)p_\infty^5
$\\
&$+\left(-\frac{3748 \nu ^2}{63}-\frac{3476 \nu}{315}-\frac{8224}{495}\right)p_\infty^6
+\left(-\frac{8885\nu ^4}{504}+\frac{4003 \nu ^3}{2156}-\frac{244914227 \nu^2}{83243160}-\frac{727820419 \nu}{16648632}+\frac{61571717}{27747720}\right) p_\infty^7
$\\
&$+\left(\frac{2714 \nu ^3}{63}+\frac{5338 \nu^2}{315}+\frac{1138058 \nu }{10395}-\frac{72192}{5005}\right)p_\infty^8
$\\
&$+\left(\frac{2197 \nu ^5}{144}+\frac{1233949 \nu^4}{388080}+\frac{92711207 \nu ^3}{21861840}+\frac{1322390051 \nu^2}{98378280}-\frac{28305965071 \nu}{2164322160}+\frac{585830087}{4088164080}\right)p_\infty^9$\\
&$+\left(-\frac{8885 \nu ^4}{252}-\frac{667 \nu^3}{30}-\frac{2641417 \nu ^2}{41580}-\frac{507193651 \nu}{3783780}+\frac{1722496}{63063}\right) p_\infty^{10}
$\\
&$+\left(-\frac{2617 \nu ^6}{192}-\frac{138454529 \nu^5}{20180160}-\frac{5395838761 \nu^4}{865728864}-\frac{35335260241 \nu^3}{2628105480}+\frac{7583454961 \nu^2}{419298880}+\frac{9889200166661 \nu}{147173906880}+\frac{364539588233}{762628426560}\right)p_\infty^{11}$\\
&$+\left(\frac{2197\nu ^5}{72}+\frac{6719 \nu ^4}{252}+\frac{102925 \nu^3}{1848}+\frac{211434659 \nu ^2}{2522520}+\frac{212030869 \nu}{1513512}-\frac{175240192}{5360355}\right)p_\infty^{12}+O(p_\infty^{13})$\\
${\mathcal E}_{20}^{{\rm 1loop} \pi}$&$0$\\
\end{tabular}
\end{ruledtabular}
\end{table*}

For $l=3$ the situation is analogous:
\begin{widetext}
\bea
E_{33}^{\rm tree}&=& \frac{ iG^3 M^4 \nu^2}{b^3}  \sqrt{1-4\nu}\sqrt{\frac{\pi}{35}}{\mathcal E}_{33}^{\rm tree}(p_\infty,\nu)\,,\nonumber\\
E_{33}^{\rm 1loop}&=& \frac{ iG^4 M^5 \nu^2}{\pi b^4}\sqrt{1-4\nu}\sqrt{\frac{\pi}{35}}[{\mathcal E}_{33}^{{\rm 1loop}\not \pi}(p_\infty,\nu)
+i\pi {\mathcal E}_{22}^{{\rm 1loop}\pi}(p_\infty,\nu)]\,,\nonumber\\
E_{31}^{\rm tree}&=& \frac{i G^3 M^4 \nu^2}{b^3}  \sqrt{1-4\nu}\sqrt{\frac{\pi}{21}}{\mathcal E}_{31}^{\rm tree}(p_\infty,\nu)\,,\nonumber\\
E_{31}^{\rm 1loop}&=& \frac{ iG^4 M^5 \nu^2}{\pi b^4}\sqrt{1-4\nu} \sqrt{\frac{\pi}{21}}[{\mathcal E}_{31}^{{\rm 1loop}\not \pi}(p_\infty,\nu)
+i\pi {\mathcal E}_{31}^{{\rm 1loop}\pi}(p_\infty,\nu)]\,.
\eea
\end{widetext}
We recall that we use the notation $M=m_1+m_2$, $\nu=m_1m_2/M^2$, and the convention  $m_2>m_1$, so that
\beq
\sqrt{1-4\nu}=\frac{(m_2-m_1)}{M}\,.
\eeq
 
The various nonvanishing coefficients are listed in Tables \ref{tab3:Elm33} and \ref{tab4:Elm31} below.

\begin{table*}
\caption{\label{tab3:Elm33}  Octupolar $(l=3, m=3)$  GW energy multipolar  coefficients at order $O(G^5)$.}
\begin{ruledtabular}
\begin{tabular}{ll}
${\mathcal E}_{33}^{\rm tree}$ & $-\frac{79}{288} p_\infty^2 +\left(\frac{102367}{16896}-\frac{22049 \nu }{6336}\right) p_\infty^4+\left(\frac{455329 \nu ^2}{164736}+\frac{1030027 \nu }{658944}-\frac{683435}{109824}\right) p_\infty^6$\\
&$+\left(-\frac{734681 \nu ^3}{329472}-\frac{7620739 \nu ^2}{5271552}-\frac{5714737 \nu}{878592}+\frac{54852269}{7028736}\right) p_\infty^8
$\\
&$+\left(\frac{168230203 \nu ^4}{89616384}+\frac{42802757 \nu ^3}{29872128}+\frac{448712695 \nu ^2}{89616384}+\frac{309695193 \nu}{39829504}-\frac{2934130291}{358465536}\right) p_\infty^{10}$\\
&$+\left(-\frac{5564951687 \nu ^5}{3405422592}-\frac{41418165979 \nu ^4}{27243380736}-\frac{9293119159 \nu^3}{2270281728}-\frac{118338907 \nu ^2}{17199104}-\frac{27118578821 \nu }{3405422592}+\frac{1750115473169}{217947045888}\right) p_\infty^{12}$\\
&$+\left(\frac{2482348837 \nu^6}{1702711296}+\frac{44812529759 \nu ^5}{27243380736}+\frac{50106874505 \nu ^4}{13621690368}+\frac{1479227003 \nu ^3}{238977024}+\frac{252469723 \nu ^2}{32587776}+\frac{1686449091391\nu }{217947045888}-\frac{833822976539}{108973522944}\right) p_\infty^{14}
$\\
&$+O(p_\infty^{16})$\\
\hline
${\mathcal E}_{33}^{{\rm 1loop}\not \pi}$ &$-\frac{244}{45} +\left(\frac{261356}{2079}-\frac{125206 \nu}{1485}\right) p_\infty^2-\frac{488}{45} p_\infty^3+\left(\frac{1923461 \nu ^2}{38610}+\frac{2086079 \nu }{24570}-\frac{5826728}{38115}\right)p_\infty^4$\\
&$+\left(\frac{355244}{1485}-\frac{250412 \nu }{1485}\right) p_\infty^5+\left(-\frac{2866843 \nu ^3}{77220}-\frac{103823 \nu
   ^2}{4004}-\frac{2357154809 \nu }{15459444}+\frac{1810373296}{19324305}\right) p_\infty^6$\\
&$+\left(\frac{1923461 \nu ^2}{19305}+\frac{19296703 \nu }{135135}-\frac{200843}{693}\right) p_\infty^7+\left(\frac{320826853 \nu ^4}{10501920}+\frac{670227731 \nu ^3}{36756720}+\frac{1552425626747
   \nu ^2}{31537265760}+\frac{1016267985743 \nu }{4505323680}-\frac{41319124064}{985539555}\right) p_\infty^8$\\
&$+\left(-\frac{2866843 \nu ^3}{38610}-\frac{8224 \nu ^2}{99}-\frac{94319879 \nu }{270270}+\frac{35112851}{90090}\right) p_\infty^9
$\\
&$+\left(-\frac{10565370611 \nu ^5}{399072960}-\frac{13356630751 \nu
   ^4}{698377680}-\frac{5217035032481 \nu ^3}{171202299840}-\frac{301376848094513 \nu ^2}{4074614736192}-\frac{2761578010121383 \nu
   }{20373073680960}-\frac{1490398757888}{15158536965}\right) p_\infty^{10}$\\
&$+\left(\frac{320826853 \nu ^4}{5250960}+\frac{1315409861 \nu ^3}{18378360}+\frac{3059342257 \nu
   ^2}{15752880}+\frac{51272769617 \nu }{110270160}-\frac{9803526947}{22054032}\right) p_\infty^{11}
$\\
&$+\left(\frac{1988177393 \nu ^6}{84015360}+\frac{81373306727 \nu ^5}{3724680960}+\frac{971017682237 \nu
   ^4}{35508625152}+\frac{245583164564833 \nu ^3}{5093268420240}+\frac{3802322267392739 \nu ^2}{309670719950592}-\frac{14631739772579291 \nu
   }{1548353599752960}+\frac{123523777898240}{403217083269}\right) p_\infty^{12}$\\
&$+O(p_\infty^{13})$\\
${\mathcal E}_{33}^{{\rm 1loop} \pi}$&$
-\frac{217}{320}p_\infty^3+\left(\frac{85235663}{3548160}-\frac{2765 \nu }{88}\right) p_\infty^5
+\left(\frac{1343181 \nu ^2}{146432}+\frac{5425989263 \nu }{73801728}-\frac{5294017}{135168}\right) p_\infty^7$\\
&$+\left(-\frac{275177 \nu ^3}{73216}-\frac{1870979947 \nu
   ^2}{98402304}-\frac{40611625259 \nu }{295206912}+\frac{21635382533}{328007680}\right) p_\infty^9$\\
&$+\left(\frac{4632607 \nu ^4}{2489344}+\frac{145757923063 \nu ^3}{20074070016}+\frac{34228579933
   \nu ^2}{955908096}+\frac{3164607147029 \nu }{13382713344}-\frac{1377243855301}{12546293760}\right) p_\infty^{11}$\\
&$+O(p_\infty^{12})$\\
\end{tabular}
\end{ruledtabular}
\end{table*}
\begin{table*}
\caption{\label{tab4:Elm31}  Octupolar $(l=3, m=1)$  GW energy multipolar  coefficients at order $O(G^5)$.}
\begin{ruledtabular}
\begin{tabular}{ll}
${\mathcal E}_{31}^{\rm tree}$ & $-\frac{653 }{240}p_\infty^2+\left(\frac{7741}{28160}-\frac{2479 \nu }{960}\right) p_\infty^4+\left(\frac{1117481 \nu ^2}{549120}+\frac{403807 \nu }{1098240}-\frac{9391}{9152}\right) p_\infty^6$\\
&$+\left(-\frac{21845 \nu ^3}{13728}-\frac{369577 \nu ^2}{675840}-\frac{432597 \nu
   }{133120}+\frac{26721867}{11714560}\right) p_\infty^8$\\
&$+\left(\frac{4868185 \nu ^4}{3734016}+\frac{7109797 \nu ^3}{9957376}+\frac{261931 \nu ^2}{106080}+\frac{801253281 \nu
   }{199147520}-\frac{18634231}{6789120}\right) p_\infty^{10}$\\
&$+\left(-\frac{30165487 \nu ^5}{27156480}-\frac{7788582821 \nu ^4}{9081126912}-\frac{1575484149 \nu ^3}{756760576}-\frac{6630542819 \nu
   ^2}{1891901440}-\frac{2955035077 \nu }{709463040}+\frac{1035005353883}{363245076480}\right) p_\infty^{12}$\\
&$+\left(\frac{22122562061 \nu ^6}{22702817280}+\frac{44550263867 \nu
   ^5}{45405634560}+\frac{8834348011 \nu ^4}{4540563456}+\frac{9694502317 \nu ^3}{3027042304}+\frac{19063203943 \nu ^2}{4779540480}+\frac{1475509165057 \nu
   }{363245076480}-\frac{92256677879}{33022279680}\right) p_\infty^{14}$\\
&$+O(p_\infty^{16})$\\
\hline
${\mathcal E}_{31}^{{\rm 1loop}\not \pi}$ &$-\frac{412}{9}+\left(\frac{216572}{17325}-\frac{33358 \nu }{495}\right) p_\infty^2 -\frac{824 }{9}p_\infty^3
+\left(\frac{466793 \nu ^2}{12870}+\frac{21712189 \nu }{450450}+\frac{276952}{17325}\right)p_\infty^4$\\
&$+\left(\frac{167308}{2475}-\frac{66716 \nu }{495}\right) p_\infty^5+\left(-\frac{659959 \nu ^3}{25740}-\frac{8619601 \nu^2}{900900}-\frac{435358009 \nu }{9909900}-\frac{53848688}{1486485}\right) p_\infty^6$\\
&$+\left(\frac{466793 \nu ^2}{6435}+\frac{2056753 \nu }{32175}-\frac{16751}{495}\right) p_\infty^7
+\left(\frac{71535049 \nu ^4}{3500640}+\frac{5489537 \nu ^3}{816816}+\frac{1683550573 \nu
   ^2}{149749600}+\frac{4374999720401 \nu }{52562109600}+\frac{8649205472}{547521975}\right) p_\infty^8$\\
&$
+\left(-\frac{659959 \nu ^3}{12870}-\frac{10095872 \nu ^2}{225225}-\frac{3212323 \nu }{21450}+\frac{30281441}{450450}\right) p_\infty^9$\\
&$+\left(-\frac{2307361967 \nu ^5}{133024320}-\frac{405194479 \nu
   ^4}{46558512}-\frac{2960271489587 \nu ^3}{399472032960}-\frac{43251988038509 \nu ^2}{1997360164800}-\frac{1058370471563 \nu }{25939742400}-\frac{12926927872}{480134655}\right)
   p_\infty^{10}$\\
&$+\left(\frac{71535049 \nu ^4}{1750320}+\frac{5717485 \nu ^3}{136136}+\frac{1916952439 \nu^2}{20420400}+\frac{926141087 \nu }{4712400}-\frac{5539960301}{61261200}\right) p_\infty^{11}$\\
&$+\left(\frac{8132078111 \nu ^6}{532097280}+\frac{210380293181 \nu ^5}{18623404800}+\frac{4627691982509 \nu ^4}{532629377280}+\frac{3429954536963 \nu
   ^3}{242536591440}-\frac{500315808331457 \nu ^2}{45273497068800}-\frac{2983187929265857 \nu }{135820491206400}+\frac{23006079337216}{530548793775}\right)
   p_\infty^{12}$\\
&$+O(p_\infty^{14})$\\
${\mathcal E}_{31}^{{\rm 1loop} \pi}$&$-\frac{8387 p_\infty^3}{1600}+\left(\frac{3043141}{1689600}-\frac{137141 \nu }{14080}\right) p_\infty^5
+\left(\frac{2014161 \nu ^2}{732160}+\frac{128620637 \nu }{9461760}-\frac{9534211}{7687680}\right) p_\infty^7$\\
&$
+\left(-\frac{80303 \nu ^3}{73216}-\frac{8291495 \nu ^2}{2236416}-\frac{429886951 \nu
   }{18923520}+\frac{1869588937}{922521600}\right) p_\infty^9$\\
&$
+\left(\frac{6604451 \nu ^4}{12446720}+\frac{1428483919 \nu ^3}{955908096}+\frac{121567511363 \nu ^2}{20074070016}+\frac{7627864738099
   \nu }{200740700160}-\frac{33979568177}{12867993600}\right) p_\infty^{11}$\\
&$+O(p_\infty^{13})$\\
\end{tabular}
\end{ruledtabular}
\end{table*}

For $l=4$ we have the following decomposition:
\begin{widetext}
\bea
E_{44}^{\rm tree}&=& \frac{ G^3 M^4 \nu^2}{b^3}  \sqrt{\frac{5\pi}{14}}{\mathcal E}_{44}^{\rm tree}(p_\infty,\nu)\,,\nonumber\\
E_{44}^{\rm 1loop}&=& \frac{ G^4 M^5 \nu^2}{\pi b^4}\sqrt{\frac{5\pi}{14}}[{\mathcal E}_{44}^{{\rm 1loop}\not \pi}(p_\infty,\nu)
+i\pi {\mathcal E}_{44}^{{\rm 1loop}\pi}(p_\infty,\nu)]\,,\nonumber\\
E_{42}^{\rm tree}&=& \frac{G^3 M^4 \nu^2}{b^3}  \sqrt{\frac{\pi}{10}}{\mathcal E}_{42}^{\rm tree}(p_\infty,\nu)\,,\nonumber\\
E_{42}^{\rm 1loop}&=& \frac{ G^4 M^5 \nu^2}{\pi b^4}\sqrt{\frac{\pi}{10}}[{\mathcal E}_{42}^{{\rm 1loop}\not \pi}(p_\infty,\nu)
+i\pi {\mathcal E}_{42}^{{\rm 1loop}\pi}(p_\infty,\nu)]\,,\nonumber\\
E_{40}^{\rm tree}&=& \frac{ G^3 M^4 \nu^2}{b^3}\pi^{1/2}{\mathcal E}_{40}^{\rm tree}(p_\infty,\nu)\,,\nonumber\\
E_{40}^{\rm 1loop}&=& \frac{ G^4 M^5 \nu^2}{\pi b^4}\sqrt{\pi}[{\mathcal E}_{40}^{{\rm 1loop}\not \pi}(p_\infty,\nu)
+i\pi {\mathcal E}_{40}^{{\rm 1loop}\pi}(p_\infty,\nu)]\,,
\eea
\end{widetext}

\begin{table*}
\caption{\label{tab4:Elm44}  Hexadecapolar $(l=4, m=4)$  GW energy multipolar  coefficients at order $O(G^5)$.}
\begin{ruledtabular}
\begin{tabular}{ll}
${\mathcal E}_{44}^{\rm tree}$ & $-\frac{5 }{96}p_\infty+\left(\frac{1013 \nu }{10560}-\frac{15221}{42240}\right) p_\infty^3
+\left(\frac{30131 \nu ^2}{21120}-\frac{1049299 \nu }{183040}+\frac{1870939}{1098240}\right) p_\infty^5$\\
&$
+\left(-\frac{792361 \nu ^3}{549120}-\frac{1014769 \nu ^2}{2196480}+\frac{5523409 \nu
   }{732160}-\frac{6337931}{2928640}\right) p_\infty^7$\\
&$
+\left(\frac{17503399 \nu ^4}{13578240}+\frac{164047777 \nu ^3}{149360640}+\frac{94789103 \nu ^2}{37340160}-\frac{1957636209 \nu
   }{199147520}+\frac{772197139}{298721280}\right) p_\infty^9$\\
&$
+\left(-\frac{6462415477 \nu ^5}{5675704320}-\frac{35910947 \nu ^4}{30638080}-\frac{12321392543 \nu ^3}{3783802880}-\frac{16324810699
   \nu ^2}{5675704320}+\frac{30953743829 \nu }{2837852160}-\frac{12915908639}{4656988160}\right) p_\infty^{11}$\\
&$
+\left(\frac{2879496071 \nu ^6}{2837852160}+\frac{6784823611 \nu
   ^5}{5675704320}+\frac{45928162989 \nu ^4}{15135211520}+\frac{50268943613 \nu ^3}{11351408640}+\frac{126254528459 \nu ^2}{45405634560}-\frac{510550143713 \nu
   }{45405634560}+\frac{340470238269}{121081692160}\right) p_\infty^{13}$\\
&$+O(p_\infty^{15})$\\
\hline
${\mathcal E}_{44}^{{\rm 1loop}\not \pi}$ &$-\frac{86}{225 p_\infty} +\left(\frac{3461 \nu
   }{825}-\frac{104989}{17325}\right) p_\infty-\frac{172 p_\infty^2}{225}+\left(\frac{1747943 \nu ^2}{42900}-\frac{1202213 \nu}{9100}+\frac{24587203}{660660}\right) p_\infty^3$\\
&$+\left(\frac{6922 \nu }{825}-\frac{4636}{495}\right) p_\infty^4
+\left(-\frac{2593871 \nu ^3}{85800}-\frac{977231 \nu ^2}{16380}+\frac{31813012199 \nu }{143143000}-\frac{1587012699947}{27054027000}\right) p_\infty^5$\\
&$+\left(\frac{1747943 \nu ^2}{21450}-\frac{36858139 \nu }{150150}+\frac{16530928}{225225}\right) p_\infty^6$\\
&$+\left(\frac{274864259 \nu ^4}{11668800}+\frac{1356272413 \nu ^3}{40840800}+\frac{66256019897677 \nu ^2}{525621096000}-\frac{819213459436721 \nu
   }{3679347672000}+\frac{210312029263747}{4496980488000}\right) p_\infty^7$\\
&$+\left(-\frac{2593871 \nu ^3}{42900}-\frac{16241669 \nu ^2}{180180}+\frac{1745806627 \nu }{4504500}-\frac{122479936}{1126125}\right)p_\infty^8$\\
&$+\left(-\frac{8637857291 \nu
   ^5}{443414400}-\frac{68471386801 \nu ^4}{3103900800}-\frac{1261030835474381 \nu ^3}{19973601648000}-\frac{11566234536573799 \nu ^2}{59421464902800}+\frac{21039692307381131407 \nu
   }{113296926414672000}-\frac{25852488107168253191}{1019672337732048000}\right) p_\infty^9$\\
&$+\left(\frac{274864259 \nu^4}{5834400}+\frac{1414820039 \nu ^3}{20420400}+\frac{392943482489 \nu ^2}{1837836000}-\frac{650522922419 \nu }{1169532000}+\frac{11535234208}{80405325}\right)p_\infty^{10}$\\
&$+\left(\frac{29805623161 \nu ^6}{1773657600}+\frac{953195267 \nu ^5}{52093440}+\frac{167224062669403 \nu
   ^4}{4438578144000}+\frac{1459775795583757 \nu ^3}{15535023504000}+\frac{4597397769225294839731 \nu ^2}{25831699222545216000}\right.$\\
&$\left.-\frac{1013975373125577819179 \nu
   }{25831699222545216000}-\frac{330320973962792825011}{15499019533527129600}\right) p_\infty^{11}$\\
&$
+\left(-\frac{8637857291 \nu ^5}{221707200}-\frac{570564343 \nu ^4}{9948400}-\frac{10874440865227 \nu
   ^3}{69837768000}-\frac{12998387008979 \nu ^2}{44442216000}+\frac{47348629994201 \nu }{69837768000}-\frac{561583936}{3357585}\right) p_\infty^{12}
$\\
&$+O(p_\infty^{13})$\\
${\mathcal E}_{44}^{{\rm 1loop} \pi}$&$ \left(\frac{36687 \nu }{35200}-\frac{6063}{17600}\right) p_\infty^4 
+\left(\frac{16458393 \nu ^2}{915200}-\frac{20447023 \nu }{600600}+\frac{17382437}{2196480}\right) p_\infty^6$\\
&$
+\left(-\frac{1004661 \nu ^3}{114400}-\frac{32245213067 \nu
   ^2}{615014400}+\frac{87483523187 \nu }{1230028800}-\frac{54655913747}{3690086400}\right) p_\infty^8$\\
&$
+\left(\frac{71904567 \nu ^4}{15558400}+\frac{166219742329 \nu
   ^3}{6970163200}+\frac{198997077737 \nu ^2}{1900953600}-\frac{406550934953 \nu }{3216998400}+\frac{302418822059}{11948851200}\right) p_\infty^{10}
$\\
&$
+\left(-\frac{393423009 \nu
   ^5}{147804800}-\frac{946698489557 \nu ^4}{79459860480}-\frac{153093590807471 \nu ^3}{3178394419200}-\frac{13925273199537 \nu ^2}{75676057600}+\frac{44602470520433393 \nu
   }{209774031667200}-\frac{908516937063221}{21515285299200}\right) p_\infty^{12}
$\\
&$+O(p_\infty^{13})$\\
\end{tabular}
\end{ruledtabular}
\end{table*}

\begin{table*}
\caption{\label{tab4:Elm42}  Hexadecapolar $(l=4, m=2)$  GW energy multipolar  coefficients at order $O(G^5)$.}
\begin{ruledtabular}
\begin{tabular}{ll}
${\mathcal E}_{42}^{\rm tree}$ & $\frac{p_\infty}{42}+\left(\frac{8509 \nu }{2112}-\frac{89875}{59136}\right) p_\infty^3+\left(\frac{384413 \nu ^2}{192192}-\frac{48155 \nu }{11648}+\frac{1575653}{1537536}\right) p_\infty^5$\\
&$
+\left(-\frac{3766261 \nu ^3}{1537536}-\frac{3035173 \nu ^2}{6150144}+\frac{1625287 \nu}{292864}-\frac{12153501}{8200192}\right) p_\infty^7$\\
&$
+\left(\frac{29041657 \nu ^4}{13069056}+\frac{280171 \nu ^3}{233376}+\frac{58899241 \nu ^2}{16084992}-\frac{601802325 \nu}{69701632}+\frac{1742622601}{836419584}\right) p_\infty^9$\\
&$
+\left(-\frac{15433710293 \nu ^5}{7945986048}-\frac{15350938957 \nu ^4}{10594648064}-\frac{21450511519 \nu^3}{5297324032}-\frac{34170211835 \nu ^2}{7945986048}+\frac{40842502963 \nu }{3972993024}-\frac{29152336093}{12108169216}\right) p_\infty^{11}$\\
&$ 
+\left(\frac{13543901087 \nu^6}{7945986048}+\frac{25607799517 \nu ^5}{15891972096}+\frac{79217206227 \nu ^4}{21189296128}+\frac{627136403 \nu ^3}{111132672}+\frac{67822062439 \nu^2}{15891972096}-\frac{1394530681127 \nu }{127135776768}+\frac{429893641017}{169514369024}\right) p_\infty^{13}$\\
&$+O(p_\infty^{14})$\\
\hline
${\mathcal E}_{42}^{{\rm 1loop}\not \pi}$ &$ \frac{20}{63 p_\infty}+\left(\frac{30438 \nu }{385}-\frac{716962}{24255}\right) p_\infty+\frac{40 p_\infty^2}{63}
+\left(\frac{2270131 \nu ^2}{30030}-\frac{26814313 \nu }{210210}+\frac{18217523}{630630}\right)p_\infty^3$\\
&$+\left(\frac{60876 \nu}{385}-\frac{40792}{693}\right) p_\infty^4+\left(-\frac{59693 \nu ^3}{1092}-\frac{265769 \nu^2}{2574}+\frac{7991559911 \nu }{69369300}-\frac{2130108181}{208107900}\right) p_\infty^5$\\
&$+\left(\frac{2270131 \nu ^2}{15015}-\frac{4180721 \nu }{15015}+\frac{430816}{6435}\right) p_\infty^6$\\
&$+\left(\frac{111197941 \nu ^4}{2722720}+\frac{458145899 \nu ^3}{9529520}+\frac{16483761765307 \nu^2}{122644922400}-\frac{1053341724839 \nu }{17520703200}-\frac{29767531327823}{4047282439200}\right) p_\infty^7$\\
&$
+\left(-\frac{59693 \nu ^3}{546}-\frac{9882731 \nu ^2}{90090}+\frac{125962313 \nu }{450450}-\frac{9188096}{143325}\right) p_\infty^8$\\
&$+\left(-\frac{1447215697 \nu ^5}{44341440}-\frac{964782389 \nu ^4}{33426624}-\frac{50125500557809 \nu ^3}{822442420800}-\frac{4950207251515 \nu^2}{23302535256}+\frac{129853429001522407 \nu }{1999357524964800}+\frac{58880721929183563}{5998072574894400}\right) p_\infty^9$\\
&$
+\left(\frac{111197941 \nu ^4}{1361360}+\frac{37794065 \nu ^3}{408408}+\frac{14630732057 \nu^2}{61261200}-\frac{166090080337 \nu }{428828400}+\frac{88473920}{1072071}\right) p_\infty^{10}$\\
&$+\left(\frac{4865124331 \nu ^6}{177365760}+\frac{609674671477 \nu
   ^5}{26072766720}+\frac{317320375195877 \nu ^4}{9321014102400}+\frac{783443732745437 \nu ^3}{8803179985600}+\frac{25464968700619766051 \nu
   ^2}{135956311697606400}+\frac{681793606976313061 \nu }{135956311697606400}-\frac{1942859690331337763}{81573787018563840}\right) p_\infty^{11}$\\
&$+\left(-\frac{1447215697 \nu ^5}{22170720}-\frac{1052265077 \nu^4}{12932920}-\frac{179907572359 \nu ^3}{997682400}-\frac{1723417248379 \nu ^2}{5431826400}+\frac{371523319571 \nu }{775975200}-\frac{30256157312}{305540235}\right)p_\infty^{12}
$\\
&$+O(p_\infty^{13})$\\
${\mathcal E}_{42}^{{\rm 1loop} \pi}$&$
\left(\frac{383301 \nu }{24640}-\frac{55809}{12320}\right) p_\infty^4
+\left(\frac{25371483 \nu ^2}{1281280}-\frac{2383869889 \nu }{107627520}+\frac{4087289}{946176}\right) p_\infty^6$\\
&$+\left(-\frac{564399 \nu ^3}{58240}-\frac{374054943 \nu
   ^2}{10250240}+\frac{25996540111 \nu }{861020160}-\frac{96329537}{20180160}\right) p_\infty^8$\\
&$
+\left(\frac{109397817 \nu ^4}{21781760}+\frac{168574288679 \nu ^3}{9758228480}+\frac{7522645440103
   \nu ^2}{117098741760}-\frac{5490850167773 \nu }{117098741760}+\frac{4897179124543}{702592450560}\right) p_\infty^{10}$\\
&$
+\left(-\frac{21013383 \nu ^5}{7390240}-\frac{1000676383981 \nu
   ^4}{111243804672}-\frac{133784528395051 \nu ^3}{4449752186880}-\frac{103990456575529 \nu ^2}{959750471680}+\frac{3035347536001489 \nu
   }{41954806333440}-\frac{6072455420594767}{587367288668160}\right) p_\infty^{12}$\\
&$+O(p_\infty^{14})$\\
\end{tabular}
\end{ruledtabular}
\end{table*}

\begin{table*}
\caption{\label{tab4:Elm40}  Hexadecapolar $(l=4, m=0)$  GW energy multipolar  coefficients at order $O(G^5)$.}
\begin{ruledtabular}
\begin{tabular}{ll}
${\mathcal E}_{40}^{\rm tree}$ & $\frac{17 }{560}p_\infty
+\left(\frac{19843 \nu }{12320}-\frac{1929}{4480}\right) p_\infty^3
+\left(\frac{761 \nu ^2}{1280}-\frac{1839207 \nu }{2562560}+\frac{670389}{2562560}\right) p_\infty^5$\\
&$
+\left(-\frac{396457 \nu ^3}{512512}-\frac{1503581 \nu ^2}{10250240}+\frac{2462583 \nu}{2050048}-\frac{3329631}{8200192}\right) p_\infty^7$\\
&$
+\left(\frac{1116599 \nu ^4}{1584128}+\frac{807407 \nu ^3}{2489344}+\frac{381886493 \nu ^2}{348508160}-\frac{34502457 \nu}{15841280}+\frac{814338753}{1394032640}\right) p_\infty^9$\\
&$+\left(-\frac{156487549 \nu ^5}{254679040}-\frac{1088172375 \nu ^4}{2648662016}-\frac{190948515 \nu ^3}{165541376}-\frac{34255168859
   \nu ^2}{26486620160}+\frac{2240946733 \nu }{827706880}-\frac{224427923}{331082752}\right) p_\infty^{11}$\\
&$+\left(\frac{2188193257 \nu ^6}{4074864640}+\frac{50040922511 \nu
   ^5}{105946480640}+\frac{2044222797 \nu ^4}{1926299648}+\frac{34409889341 \nu ^3}{21189296128}+\frac{273324627217 \nu ^2}{211892961280}-\frac{2492229588049 \nu
   }{847571845120}+\frac{10998358955}{15410397184}\right) p_\infty^{13}$\\
&$+O(p_\infty^{15})$\\
\hline
${\mathcal E}_{40}^{{\rm 1loop}\not \pi}$ &$\frac{134}{315 p_\infty}+\left(\frac{3289 \nu}{105}-\frac{138529}{17325}\right) p_\infty+\frac{268 p_\infty^2}{315}
+\left(\frac{1467553 \nu ^2}{60060}-\frac{4027761 \nu }{140140}+\frac{49578149}{6306300}\right)p_\infty^3$\\
&$+\left(\frac{6578 \nu }{105}-\frac{57748}{3465}\right) p_\infty^4
+\left(-\frac{703123 \nu ^3}{40040}-\frac{382433 \nu^2}{11700}+\frac{179135623 \nu }{12612600}-\frac{1882441}{491400}\right) p_\infty^5$\\
&$+\left(\frac{1467553 \nu ^2}{30030}-\frac{3486141 \nu }{50050}+\frac{614608}{32175}\right) p_\infty^6$\\
&$+\left(\frac{226771 \nu ^4}{17472}+\frac{1369015693 \nu ^3}{95295200}+\frac{38712094381 \nu^2}{1109908800}-\frac{6676951147 \nu }{3773689920}-\frac{53758127019}{23061438400}\right) p_\infty^7$\\
&$+\left(-\frac{703123 \nu ^3}{20020}-\frac{27589901 \nu ^2}{900900}+\frac{4768409 \nu }{81900}-\frac{3812416}{225225}\right) p_\infty^8$\\
&$+\left(-\frac{70106983 \nu^5}{6821760}-\frac{36143146951 \nu ^4}{4345461120}-\frac{15589474095121 \nu ^3}{1035668233600}-\frac{66561051920257 \nu ^2}{1165126762800}+\frac{1633662093384023 \nu
   }{148100557404800}+\frac{6195358708777957}{2399229029957760}\right) p_\infty^9$\\
&$
+\left(\frac{226771 \nu ^4}{8736}+\frac{26085447 \nu ^3}{972400}+\frac{467945747 \nu^2}{7207200}-\frac{14405864047 \nu }{171531360}+\frac{51710752}{2436525}\right) p_\infty^{10}$\\
&$+\left(\frac{1013850819 \nu ^6}{118243840}+\frac{1744750337587 \nu ^5}{260727667200}+\frac{151247381071783 \nu
   ^4}{18642028204800}+\frac{625304645751557 \nu ^3}{27963042307200}+\frac{29752346934860659 \nu ^2}{592402229619200}-\frac{44037687359619017 \nu
   }{15994860199718400}\right. $\\
&$\left.-\frac{39092903516829427}{6854940085593600}\right) p_\infty^{11}$\\
&$+\left(-\frac{70106983 \nu ^5}{3410880}-\frac{170635139 \nu ^4}{7054320}-\frac{1189194403\nu ^3}{23876160}-\frac{3408477833 \nu ^2}{39793600}+\frac{1169239870439 \nu }{10863652800}-\frac{389433152}{15431325}\right) p_\infty^{12}$\\
&$+O(p_\infty^{13})$\\
${\mathcal E}_{40}^{{\rm 1loop} \pi}$&$0$\\
\end{tabular}
\end{ruledtabular}
\end{table*}

\clearpage

Finally, for $l=5$ we have the following decomposition:
\begin{widetext}
\bea
E_{55}^{\rm tree}&=& \frac{ G^3 M^4 \nu^2}{b^3}i \sqrt{1-4\nu}\sqrt{\frac{7\pi}{11}}{\mathcal E}_{55}^{\rm tree}(p_\infty,\nu)\,,\nonumber\\
E_{55}^{\rm 1loop}&=& \frac{ G^4 M^5 \nu^2}{\pi b^4}i \sqrt{1-4\nu}\sqrt{\frac{7\pi}{11}} [{\mathcal E}_{55}^{{\rm 1loop}\not \pi}(p_\infty,\nu)
+i\pi {\mathcal E}_{55}^{{\rm 1loop}\pi}(p_\infty,\nu)]\,,\nonumber\\
E_{53}^{\rm tree}&=& \frac{ G^3 M^4 \nu^2}{b^3}  i \sqrt{1-4\nu} \sqrt{\frac{7\pi}{55}}{\mathcal E}_{53}^{\rm tree}(p_\infty,\nu)\,,\nonumber\\
E_{53}^{\rm 1loop}&=& \frac{ G^4 M^5 \nu^2}{\pi b^4}i \sqrt{1-4\nu}\sqrt{\frac{7\pi}{55}}[{\mathcal E}_{53}^{{\rm 1loop}\not \pi}(p_\infty,\nu)
+i\pi {\mathcal E}_{53}^{{\rm 1loop}\pi}(p_\infty,\nu)]\,,\nonumber\\
E_{51}^{\rm tree}&=& \frac{ G^3 M^4 \nu^2}{b^3}i  \sqrt{1-4\nu}\sqrt{\frac{\pi}{330}}{\mathcal E}_{51}^{\rm tree}(p_\infty,\nu)\,,\nonumber\\
E_{51}^{\rm 1loop}&=&\frac{G^4 M^5 \nu^2}{\pi b^4}i  \sqrt{1-4\nu}\sqrt{\frac{\pi}{330}}[{\mathcal E}_{51}^{{\rm 1loop}\not \pi}(p_\infty,\nu)
+i\pi {\mathcal E}_{51}^{{\rm 1loop}\pi}(p_\infty,\nu)]\,,
\eea
\end{widetext}

\begin{table*}
\caption{\label{tab4:Elm55}  Dotriacontapolar $(l=5, m=5)$  GW energy multipolar  coefficients at order $O(G^5)$.}
\begin{ruledtabular}
\begin{tabular}{ll}
${\mathcal E}_{55}^{\rm tree}$ & $ \frac{25}{384}p_\infty^2+\left(\frac{193 \nu }{9984}+\frac{5165}{13312}\right) p_\infty^4
+\left(-\frac{36811 \nu ^2}{99840}+\frac{28799 \nu }{10920}-\frac{167107}{133120}\right) p_\infty^6$\\
&$
+\left(\frac{1530067 \nu ^3}{3394560}-\frac{273031 \nu ^2}{1118208}-\frac{27427147 \nu}{6336512}+\frac{17156177}{9748480}\right) p_\infty^8$\\
&$+\left(-\frac{463068001 \nu ^4}{1031946240}-\frac{958131179 \nu ^3}{2407874560}+\frac{87511483 \nu ^2}{1111326720}+\frac{200396887 \nu
   }{34398208}-\frac{2200615583}{1031946240}\right) p_\infty^{10}$\\
&$+\left(\frac{174482779 \nu ^5}{412778496}+\frac{33365629067 \nu ^4}{57788989440}+\frac{5168025337 \nu
   ^3}{4815749120}-\frac{366531323 \nu ^2}{1481768960}-\frac{78264994763 \nu }{11557797888}+\frac{12794822987}{5438963712}\right) p_\infty^{12}$\\
&$+\left(-\frac{3720492281 \nu
   ^6}{9493905408}-\frac{168182221579 \nu ^5}{265829351424}-\frac{1811225807977 \nu ^4}{1329146757120}-\frac{12188403971 \nu ^3}{7911587840}+\frac{1325098524917 \nu
   ^2}{2658293514240}+\frac{11071833887687 \nu }{1519024865280}-\frac{26136414484787}{10633174056960}\right) p_\infty^{14}$\\
&$+O(p_\infty^{15})$\\
\hline
${\mathcal E}_{55}^{{\rm 1loop}\not \pi}$ &$
\frac{172}{315}+\left(\frac{39862}{5733}-\frac{1174 \nu }{819}\right) p_\infty^2+\frac{344}{315} p_\infty^3+\left(-\frac{97091 \nu ^2}{8190}+\frac{1268633 \nu }{19110}-\frac{202236626}{6831825}\right)p_\infty^4$\\
&$+\left(\frac{61160}{5733}-\frac{2348 \nu }{819}\right) p_\infty^5
+\left(\frac{3079469 \nu
   ^3}{278460}+\frac{87240037 \nu ^2}{4873050}-\frac{59725765943 \nu }{418107690}+\frac{380105618084}{7316884575}\right) p_\infty^6$\\
&$+\left(-\frac{97091 \nu ^2}{4095}+\frac{231979 \nu }{1911}-\frac{2743009}{47775}\right) p_\infty^7$\\
&$+\left(-\frac{56641951 \nu
   ^4}{6046560}-\frac{940476863 \nu ^3}{49380240}-\frac{3182492880631 \nu ^2}{70613743200}+\frac{4505943827874031 \nu }{25209106322400}-\frac{50694426469063864}{1013878744904025}\right)p_\infty^8$\\
&$+\left(\frac{3079469 \nu ^3}{139230}+\frac{24668179 \nu ^2}{974610}-\frac{1748485712 \nu }{7309575}+\frac{4862336624}{51167025}\right) p_\infty^9$\\
&$
+\left(\frac{676928069 \nu ^5}{84651840}+\frac{720193919 \nu ^4}{47028800}+\frac{10264244283899 \nu ^3}{254209475520}+\frac{620085931039387 \nu
   ^2}{7562731896720}-\frac{457499202399816307 \nu }{2590076793704400}+\frac{14488632122202016}{428082136737255}\right) p_\infty^{10}$\\
&$+\left(-\frac{56641951 \nu ^4}{3023280}-\frac{4230189361 \nu
   ^3}{123450600}-\frac{14792220551 \nu ^2}{246901200}+\frac{36938383957 \nu }{101665200}-\frac{153894611263}{1196521200}\right) p_\infty^{11}$\\
&$+\left(-\frac{10860129095 \nu
   ^6}{1557593856}-\frac{3475338568759 \nu ^5}{272578924800}-\frac{684730388020243 \nu ^4}{23387271747840}-\frac{465135135416814979 \nu ^3}{6957713344982400}-\frac{121775306192812796501
   \nu ^2}{1246424647801132800}\right.$\\
&$\left. +\frac{11168526282294669730081 \nu }{101485206849913286400}+\frac{31465667176471507264}{7532105195892001725}\right)
   p_\infty^{12}
$\\
&$+O(p_\infty^{13})$\\
${\mathcal E}_{55}^{{\rm 1loop} \pi}$&$\frac{7}{512} p_\infty^3+\left(\frac{493719}{931840}-\frac{40361 \nu }{66560}\right) p_\infty^5 
+\left(-\frac{1579161 \nu ^2}{266240}+\frac{1015938723 \nu }{52183040}-\frac{1003700731}{156549120}\right) p_\infty^7$\\
&$
+\left(\frac{9471427 \nu ^3}{2263040}+\frac{6034311845 \nu
   ^2}{304152576}-\frac{1838772599 \nu }{39137280}+\frac{47070987883}{3548446720}\right) p_\infty^9$\\
&$
+\left(-\frac{232710121 \nu ^4}{85995520}-\frac{2301309727319 \nu
   ^3}{161809170432}-\frac{216268372685 \nu ^2}{5056536576}+\frac{9354220372743703 \nu }{106794052485120}-\frac{1246630718918197}{53397026242560}\right)
   p_\infty^{11}$\\
&$+O(p_\infty^{13})$\\
\end{tabular}
\end{ruledtabular}
\end{table*}
 
\begin{table*}
\caption{\label{tab4:Elm53}  Dotriacontapolar $(l=5, m=3)$ non-linear memory coefficients at order $O(G^5)$.}
\begin{ruledtabular}
\begin{tabular}{ll}
${\mathcal E}_{53}^{\rm tree}$ & $-\frac{7}{1152}p_\infty^2+\left(\frac{318445}{279552}-\frac{47581 \nu }{29952}\right) p_\infty^4
 +\left(-\frac{14935 \nu ^2}{59904}+\frac{186053 \nu }{59904}-\frac{742793}{559104}\right) p_\infty^6$\\
&$
+\left(\frac{47999 \nu ^3}{63648}-\frac{7847813 \nu ^2}{28514304}-\frac{12787199 \nu
   }{2715648}+\frac{67840813}{38019072}\right) p_\infty^8$\\
&$
+\left(-\frac{527190301 \nu ^4}{619167744}-\frac{856492163 \nu ^3}{1444724736}-\frac{3516444317 \nu ^2}{8668348416}+\frac{3217311589 \nu
   }{481574912}-\frac{9812242103}{4334174208}\right) p_\infty^{10}$\\
&$
+\left(\frac{1024949429 \nu ^5}{1238335488}+\frac{1640303159 \nu ^4}{1824915456}+\frac{5035024601 \nu
   ^3}{2889449472}+\frac{1560805429 \nu ^2}{3852599296}-\frac{21458308927 \nu }{2667184128}+\frac{717443527817}{277387149312}\right) p_\infty^{12}$\\
&$
+\left(-\frac{1689180133 \nu
   ^6}{2190901248}-\frac{823275887585 \nu ^5}{797488054272}-\frac{1656694164877 \nu ^4}{797488054272}-\frac{18044578607 \nu ^3}{6995509248}-\frac{275773487695 \nu
   ^2}{1594976108544}+\frac{56447324034611 \nu }{6379904434176}-\frac{17634679751231}{6379904434176}\right) p_\infty^{14}$
\\
&$+O(p_\infty^{16})$\\
\hline
${\mathcal E}_{53}^{{\rm 1loop}\not \pi}$ &$-\frac{76}{315}+\left(\frac{301706}{12285}-\frac{444422 \nu }{12285}\right) p_\infty^2-\frac{152}{315} p_\infty^3
+\left(-\frac{530111 \nu ^2}{24570}+\frac{17253191 \nu}{171990}-\frac{59613506}{1576575}\right) p_\infty^4$\\
&$+\left(\frac{584536}{12285}-\frac{888844 \nu }{12285}\right)p_\infty^5+\left(\frac{19028081 \nu ^3}{835380}+\frac{286421 \nu ^2}{6630}-\frac{299350090861 \nu }{2090538450}+\frac{43226262796}{1463376915}\right) p_\infty^6$\\
&$+\left(-\frac{530111 \nu ^2}{12285}+\frac{16515209 \nu }{85995}-\frac{10436437}{143325}\right) p_\infty^7$\\
&$+\left(-\frac{349087051 \nu ^4}{18139680}-\frac{1293471077 \nu ^3}{34186320}-\frac{141916798747483 \nu ^2}{1906571066400}+\frac{364292214731149 \nu
   }{2669199492960}-\frac{51522446112392}{3508230951225}\right) p_\infty^8$\\
&$+\left(\frac{19028081 \nu ^3}{417690}+\frac{5309123 \nu ^2}{139230}-\frac{1933873112 \nu }{7309575}+\frac{1487204648}{17055675}\right)
   p_\infty^9$\\
&$+\left(\frac{584115983 \nu ^5}{36279360}+\frac{770320865 \nu ^4}{27349056}+\frac{226715397193171 \nu
   ^3}{3813142132800}+\frac{14732169371986799 \nu ^2}{113440978450800}-\frac{302186202346050563 \nu }{2317437131209200}+\frac{206036079523168}{67591916326935}\right)
   p_\infty^{10}$\\
&$
+\left(-\frac{349087051 \nu
   ^4}{9069840}-\frac{12524347427 \nu ^3}{222211080}-\frac{602919299051 \nu ^2}{6666332400}+\frac{16930875706193 \nu }{46664326800}-\frac{5024872138159}{46664326800}\right)p_\infty^{11}$\\
&$+\left(-\frac{320973072991 \nu ^6}{23363907840}-\frac{244686946697 \nu ^5}{10903156992}-\frac{538410831572561 \nu ^4}{12992928748800}-\frac{174787534239373357 \nu
   ^3}{1897558184995200}-\frac{186730761343165366259 \nu ^2}{1246424647801132800}\right.$\\
&$\left.+\frac{149737045321626692673413 \nu
   }{1928218930148352441600}+\frac{1902555075824956864}{109160944868000025}\right) p_\infty^{12}$\\
&$+O(p_\infty^{13})$\\
${\mathcal E}_{53}^{{\rm 1loop} \pi}$&$-\frac{19}{512} p_\infty^3+\left(\frac{34630391}{8386560}-\frac{628711 \nu }{66560}\right) p_\infty^5
+\left(-\frac{2164641 \nu ^2}{266240}+\frac{5005509911 \nu }{234823680}-\frac{11714557}{2007040}\right) p_\infty^7$\\
&$
+\left(\frac{14155847 \nu ^3}{2263040}+\frac{6075164527 \nu
   ^2}{304152576}-\frac{552293524297 \nu }{15968010240}+\frac{82129233059}{10645340160}\right) p_\infty^9$\\
&$
+\left(-\frac{351506561 \nu ^4}{85995520}-\frac{7258999847597 \nu
   ^3}{485427511296}-\frac{786204017017 \nu ^2}{20226146304}+\frac{34587654940483 \nu }{624526622720}-\frac{121779868912411}{10679405248512}\right)
   p_\infty^{11}$\\
&$+O(p_\infty^{13})$\\
\end{tabular}
\end{ruledtabular}
\end{table*}

\begin{table*}
\caption{\label{tab4:Elm51}  Dotriacontapolar $(l=5, m=1)$ non-linear memory coefficients at order $O(G^5)$.}
\begin{ruledtabular}
\begin{tabular}{ll}
${\mathcal E}_{51}^{\rm tree}$ & $ -\frac{59}{96} p_\infty^2+\left(\frac{287335}{46592}-\frac{5593 \nu }{384}\right) p_\infty^4
+\left(-\frac{6869 \nu ^2}{19968}+\frac{3701309 \nu }{279552}-\frac{1260535}{186368}\right) p_\infty^6$\\
&$
+\left(\frac{1532821 \nu ^3}{339456}-\frac{3654887 \nu ^2}{4752384}-\frac{66794487 \nu}{3168256}+\frac{60956463}{6336512}\right) p_\infty^8$\\
&$
+\left(-\frac{139682059 \nu ^4}{25798656}-\frac{105238765 \nu ^3}{30098432}-\frac{1446372575 \nu ^2}{361181184}+\frac{14515887 \nu
   }{442624}-\frac{1039811}{82992}\right) p_\infty^{10}$\\
&$
+\left(\frac{1101104561 \nu ^5}{206389248}+\frac{4328999489 \nu ^4}{825556992}+\frac{710261739 \nu ^3}{68796416}+\frac{8368866665 \nu
   ^2}{1926299648}-\frac{238123064089 \nu }{5778898944}+\frac{667333469723}{46231191552}\right) p_\infty^{12}$\\
&$
+\left(-\frac{94570468519 \nu ^6}{18987810816}-\frac{232859344645 \nu
   ^5}{37975621632}-\frac{902735721259 \nu ^4}{75951243264}-\frac{2722011453179 \nu ^3}{177219567616}-\frac{197872447847 \nu ^2}{62548082688}+\frac{98456394167059 \nu
   }{2126634811392}-\frac{722343279437}{46739226624}\right) p_\infty^{14}$
\\
&$+O(p_\infty^{15})$\\
\hline
${\mathcal E}_{51}^{{\rm 1loop}\not \pi}$ &$-\frac{88}{9}+\left(\frac{530708}{4095}-\frac{189236 \nu }{585}\right) p_\infty^2
-\frac{176 }{9}p_\infty^3+\left(-\frac{78673 \nu ^2}{585}+\frac{2114201 \nu }{4095}-\frac{28419172}{143325}\right)p_\infty^4$\\
&$+\left(\frac{1091056}{4095}-\frac{378472 \nu }{585}\right) p_\infty^5+\left(\frac{175247 \nu ^3}{1170}+\frac{607565 \nu^2}{1989}-\frac{854819837 \nu }{1576575}+\frac{533022968}{3216213}\right) p_\infty^6$\\
&$+\left(-\frac{157346 \nu ^2}{585}+\frac{4141294 \nu }{4095}-\frac{8099522}{20475}\right) p_\infty^7$\\
&$+\left(-\frac{382376483 \nu ^4}{3023280}-\frac{123618659 \nu ^3}{503880}-\frac{5112002872553
   \nu ^2}{11768957200}+\frac{30427242302851 \nu }{63552368880}-\frac{211516292668528}{2839996484325}\right) p_\infty^8$\\
&$+\left(\frac{175247 \nu ^3}{585}+\frac{15701561 \nu ^2}{69615}-\frac{207176 \nu }{175}+\frac{1095164008}{2436525}\right) p_\infty^9$\\
&$
+\left(\frac{48844387 \nu ^5}{465120}+\frac{438528887 \nu
   ^4}{2489760}+\frac{10896261330233 \nu ^3}{33448615200}+\frac{10608522815417 \nu ^2}{14443720200}-\frac{4001760438898091 \nu }{7573323958200}+\frac{15159097840064}{567999296865}\right)p_\infty^{10}$\\
&$+\left(-\frac{382376483 \nu ^4}{1511640}-\frac{196507541 \nu ^3}{587860}-\frac{9079279229 \nu
   ^2}{17635800}+\frac{34636111813 \nu }{21785400}-\frac{200099012843}{370351800}\right) p_\infty^{11}$\\
&$+\left(-\frac{49425013169 \nu ^6}{556283520}-\frac{28173782179 \nu ^5}{204946560}-\frac{618700698191959 \nu ^4}{2784199017600}-\frac{242848296252849257 \nu
   ^3}{496979524641600}-\frac{9121377580725781817 \nu ^2}{10933549542115200}\right.$\\
&$\left.+\frac{993355312247482041497 \nu }{2416314448807459200}+\frac{297521919538278016}{6292485543769425}\right)p_\infty^{12}$\\
&$+O(p_\infty^{13})$\\
${\mathcal E}_{51}^{{\rm 1loop} \pi}$&$-\frac{137}{256} p_\infty^3+\left(\frac{18931009}{2795520}-\frac{469843 \nu }{16640}\right) p_\infty^5
+\left(-\frac{2457381 \nu ^2}{133120}+\frac{6697393 \nu }{174720}-\frac{202290079}{19568640}\right) p_\infty^7$\\
&$
+\left(\frac{16531657 \nu ^3}{1131520}+\frac{3047003359 \nu
   ^2}{76038144}-\frac{6859284749 \nu }{126730240}+\frac{101626482011}{7984005120}\right) p_\infty^9$\\
&$
+\left(-\frac{411845581 \nu ^4}{42997760}-\frac{350075002117 \nu
   ^3}{11557797888}-\frac{75133404131 \nu ^2}{1019805696}+\frac{618844017673229 \nu }{7628146606080}-\frac{30951999593063}{1779900874752}\right)
   p_\infty^{11}$\\
&$+O(p_\infty^{12})$\\
\end{tabular}
\end{ruledtabular}
\end{table*}

We refrain from presenting explicit results here for other values of $l,m$; they are instead collected in an associated ancillary file, up to $l=17$ and at 7.5PN accuracy.  
The results are presented as \texttt{memory\_l\_m.m}, with obvious reference to the corresponding $l$ and $m$.

\subsection{Checks}

As a  check of our results, we have derived (by considering $l=0,1$) the PN-expanded energy,  and linear momentum, losses at $O(G^3)$ (tree level, see Ref. \cite{Herrmann:2021tct}) and $O(G^4)$ (1-loop level, see Ref. \cite{Dlapa:2022lmu}), in agreement with the literature. 
 We have also checked that the $\nu \to 0$ limit of our, cm-frame, nonlinear memory results agree
with the rest-frame results of Ref. \cite{Georgoudis:2025vkk}.  

\section{Gravitational-wave energy spectrum at the 1-loop level and at the 7.5PN level of accuracy}

We have also computed the radiated-energy spectrum  up to the 1-loop order and at 7.5PN accuracy,
\beq
\frac{dE^+_{\rm rad}}{d\omega}=\int d\Omega \frac{dE}{d\omega d\Omega}\,,
\eeq
integrated over the angles.
Passing to the variable $u=\frac{\omega b}{p_\infty}$,   
we write
\beq
\frac{dE_{\rm rad}}{d u}={\mathcal F}(u)={\mathcal F}^{\rm tree}(u)+{\mathcal F}^{\rm 1 loop}(u)\,,
\eeq
such that
\bea
E_{\rm rad}=\int_0^\infty du \, {\mathcal F}(u)\,,
\eea
where, as usual, the integration over $u$ runs only over positive frequencies.

The tree level energy spectrum was computed in Ref. \cite{Bini:2024ijq}
in the frame of one of the two bodies at order $O(v^{30})$. In the present paper we have chosen to work in the cm frame. Using the simple transformation indicated in Appendix B of Ref. \cite{Bini:2026dvn} (see also Ref. \cite{Bini:2024ijq}) it is straightforward to transform
the rest-frame energy spectrum into its cm-frame counterpart.
Let us indicate here the
beginning of the PN expansion of the cm-frame energy spectrum.  
\bea
{\mathcal F}^{\rm tree}(u)&=& p_\infty \underbrace{{\mathcal F}^{\rm tree}_{1}(u)}_{\rm Newtonian}+{\mathcal F}^{\rm tree}_{3}(u)p_\infty^3 +{\mathcal F}^{\rm tree}_{5}(u)p_\infty^5\nonumber\\
&+&{\mathcal F}^{\rm tree}_{7}(u)p_\infty^7 +\ldots +\underbrace{{\mathcal F}^{\rm tree}_{15}(u)}_{\rm 7.5PN}p_\infty^{15}\nonumber\\ 
&+& O(p_\infty^{16}).
\eea
The various coefficients ${\mathcal F}^{\rm tree}_{k}(u)$ are bilinear in the Bessel $K$ functions of order $0$ and $1$, namely
\bea
{\mathcal F}^{\rm tree}_{k}(u)&=& \frac{G^2  M^4  \nu^2}{\pi b^3} \left[C_{00}^{(k)} K_0(u)^2+C_{01}^{(k)} K_0(u)K_1(u)\right.\nonumber\\
&+&\left.  C_{11}^{(k)} K_1(u)^2 
\right]\,.
\eea
Explicitly, for example,
\bea
C_{00}^{(1)}&=&\frac{32 u^2 \left(3 u^2+1\right)}{15 } \,,\nonumber\\
C_{01}^{(1)}&=& \frac{96 u^3}{5 }\,,\nonumber\\
C_{11}^{(1)}&=&\frac{32 u^2 \left(u^2+1\right)}{5 } \,,
\eea
and
\bea
C_{00}^{(3)}&=&  \nu  \left(\frac{128 u^2}{105}-\frac{64 u^6}{21}\right)+\frac{16 u^6}{21}+\frac{16 u^4}{105}-\frac{160 u^2}{21}\,,\nonumber\\
C_{01}^{(3)}&=&  \nu  \left(\frac{128 u^5}{35}+\frac{512 u^3}{105}\right)-\frac{736 u^5}{105}+\frac{288 u^3}{35}\,,\nonumber\\
C_{11}^{(3)}&=&  \nu  \left(\frac{32 u^4}{105}-\frac{64 u^6}{21}\right)+\frac{16 u^6}{21}-\frac{104 u^4}{35}+\frac{160 u^2}{7}\,.\nonumber\\
\eea
The coefficients of  the tree-level spectrum up to $O(p_\infty^{15})$ are listed in Tables \ref{Cck}, \ref{Cck1} and \ref{Cck2} below.

\begin{table*}
\caption{\label{Cck}  Energy spectrum at the tree level: list of coefficients from $p_\infty^5$ up to $p_\infty^{11}$}
\begin{ruledtabular}
\begin{tabular}{ll}
$C_{00}^{(5)}$& $\nu ^2 \left(\frac{32 u^8}{63}+\frac{5576 u^6}{945}-\frac{256 u^4}{945}-\frac{128 u^2}{315}\right)+\nu  \left(-\frac{64 u^8}{189}-\frac{544 u^6}{63}+\frac{64 u^4}{27}-\frac{1472
   u^2}{315}\right)+\frac{32 u^8}{567}+\frac{7592 u^6}{2835}-\frac{608 u^4}{105}+\frac{1408 u^2}{315}$\\
$C_{01}^{(5)}$& $\nu ^2 \left(-\frac{928 u^7}{189}-\frac{1664 u^5}{315}-\frac{2048 u^3}{945}\right)+\nu  \left(\frac{128 u^7}{21}+\frac{11392 u^5}{945}-\frac{7424 u^3}{945}\right)-\frac{32
   u^7}{27}-\frac{1184 u^5}{567}-\frac{64 u^3}{45}$\\
$C_{11}^{(5)}$& $\nu ^2 \left(\frac{32 u^8}{63}+\frac{3496 u^6}{945}+\frac{32 u^4}{315}\right)+\nu  \left(-\frac{64 u^8}{189}-\frac{1088 u^6}{189}-\frac{7472 u^4}{945}\right)+\frac{32 u^8}{567}+\frac{856
   u^6}{405}+\frac{31232 u^4}{2835}+\frac{1024 u^2}{63}$\\
\hline
$C_{00}^{(7)}$& $\nu ^3 \left(-\frac{64 u^{10}}{1485}-\frac{81232 u^8}{31185}-\frac{282928 u^6}{31185}+\frac{512 u^4}{2079}+\frac{128 u^2}{693}\right)+\nu ^2 \left(\frac{16 u^{10}}{297}+\frac{17284
   u^8}{3465}+\frac{32176 u^6}{1485}-\frac{4288 u^4}{10395}+\frac{5728 u^2}{3465}\right)$\\
&$+\nu  \left(-\frac{32 u^{10}}{1485}-\frac{9304 u^8}{3465}-\frac{514904 u^6}{31185}-\frac{34784
   u^4}{10395}+\frac{2672 u^2}{693}\right)+\frac{4 u^{10}}{1485}+\frac{12337 u^8}{31185}+\frac{5954 u^6}{2079}+\frac{3776 u^4}{3465}+\frac{19072 u^2}{3465}$\\
$C_{01}^{(7)}$& $\nu ^3 \left(\frac{3328 u^9}{3465}+\frac{351088 u^7}{31185}+\frac{174848 u^5}{31185}+\frac{2560 u^3}{2079}\right)+\nu ^2 \left(-\frac{128 u^9}{77}-\frac{1556 u^7}{63}-\frac{30080
   u^5}{2079}+\frac{38272 u^3}{10395}\right)$\\
&$+\nu  \left(\frac{4640 u^9}{6237}+\frac{66928 u^7}{4455}+\frac{607216 u^5}{31185}-\frac{15808 u^3}{10395}\right)-\frac{208
   u^9}{2079}-\frac{86243 u^7}{31185}-\frac{24536 u^5}{3465}-\frac{10240 u^3}{693}$\\
$C_{11}^{(7)}$& $\nu ^3 \left(-\frac{64 u^{10}}{1485}-\frac{66928 u^8}{31185}-\frac{137776 u^6}{31185}-\frac{4192 u^4}{31185}\right)+\nu ^2 \left(\frac{16 u^{10}}{297}+\frac{43492 u^8}{10395}+\frac{116728
   u^6}{10395}+\frac{9448 u^4}{2079}\right)$\\
&$+\nu  \left(-\frac{32 u^{10}}{1485}-\frac{72472 u^8}{31185}-\frac{14968 u^6}{1485}-\frac{140036 u^4}{31185}\right)+\frac{4
   u^{10}}{1485}+\frac{10819 u^8}{31185}+\frac{10250 u^6}{6237}-\frac{26008 u^4}{10395}-\frac{19328 u^2}{3465}$\\
\hline
$C_{00}^{(9)}$& $\nu ^4 \left(\frac{128 u^{12}}{57915}+\frac{55232 u^{10}}{135135}+\frac{2637808 u^8}{405405}+\frac{4983416 u^6}{405405}-\frac{256 u^4}{1287}-\frac{128 u^2}{1287}\right)$\\
&$+\nu ^3
   \left(-\frac{256 u^{12}}{57915}-\frac{33136 u^{10}}{31185}-\frac{7171816 u^8}{405405}-\frac{152756 u^6}{4455}-\frac{192 u^4}{5005}-\frac{1024 u^2}{1287}\right)$\\
&$+\nu ^2 \left(\frac{896
   u^{12}}{289575}+\frac{358192 u^{10}}{405405}+\frac{6839792 u^8}{405405}+\frac{25425112 u^6}{675675}+\frac{20752 u^4}{12285}-\frac{10928 u^2}{6435}\right)$\\
&$+\nu  \left(-\frac{256
   u^{12}}{289575}-\frac{7676 u^{10}}{27027}-\frac{12761906 u^8}{2027025}-\frac{33688244 u^6}{2027025}-\frac{348728 u^4}{96525}+\frac{2456 u^2}{1365}\right)$\\
&$+\frac{128
   u^{12}}{1447875}+\frac{105232 u^{10}}{3378375}+\frac{2714144 u^8}{3378375}+\frac{9096704 u^6}{3378375}+\frac{106688 u^4}{32175}-\frac{119296 u^2}{45045}$\\
$C_{01}^{(9)}$& $\nu ^4 \left(-\frac{128 u^{11}}{1485}-\frac{1554944 u^9}{405405}-\frac{168032 u^7}{9009}-\frac{460160 u^5}{81081}-\frac{1024 u^3}{1287}\right)$\\
&$+\nu ^3 \left(\frac{12224
   u^{11}}{57915}+\frac{280144 u^9}{27027}+\frac{461872 u^7}{9009}+\frac{6225088 u^5}{405405}-\frac{102656 u^3}{45045}\right)$\\
&$+\nu ^2 \left(-\frac{46528 u^{11}}{289575}-\frac{2075728
   u^9}{225225}-\frac{108794552 u^7}{2027025}-\frac{789952 u^5}{32175}+\frac{92992 u^3}{135135}\right)$\\
&$+\nu  \left(\frac{14096 u^{11}}{289575}+\frac{2190268 u^9}{675675}+\frac{270620
   u^7}{11583}+\frac{196384 u^5}{11583}-\frac{341792 u^3}{61425}\right)$\\
&$-\frac{1472 u^{11}}{289575}-\frac{426656 u^9}{1126125}-\frac{10448852 u^7}{3378375}-\frac{353504
   u^5}{307125}+\frac{105088 u^3}{25025}$\\
$C_{11}^{(9)}$& $\nu ^4 \left(\frac{128 u^{12}}{57915}+\frac{148672 u^{10}}{405405}+\frac{214528 u^8}{45045}+\frac{185096 u^6}{36855}+\frac{736 u^4}{6237}\right)$\\
&$+\nu ^3 \left(-\frac{256
   u^{12}}{57915}-\frac{77776 u^{10}}{81081}-\frac{250256 u^8}{19305}-\frac{846628 u^6}{57915}-\frac{1179392 u^4}{405405}\right)$\\
&$+\nu ^2 \left(\frac{896 u^{12}}{289575}+\frac{1631248
   u^{10}}{2027025}+\frac{25636232 u^8}{2027025}+\frac{37898176 u^6}{2027025}+\frac{33716 u^4}{225225}\right)$\\
&$+\nu  \left(-\frac{256 u^{12}}{289575}-\frac{105452
   u^{10}}{405405}-\frac{9728696 u^8}{2027025}-\frac{3907912 u^6}{405405}-\frac{115378 u^4}{57915}\right)$\\
&$+\frac{128 u^{12}}{1447875}+\frac{290384 u^{10}}{10135125}+\frac{32624
   u^8}{51975}+\frac{1332664 u^6}{675675}+\frac{2550368 u^4}{482625}+\frac{130048 u^2}{45045}$\\
\hline
$C_{00}^{(11)}$& $\nu ^5 \left(-\frac{256 u^{14}}{3378375}-\frac{1004032 u^{12}}{30405375}-\frac{1829104 u^{10}}{1216215}-\frac{1921856 u^8}{155925}-\frac{1728352 u^6}{111375}+\frac{1024
   u^4}{6435}+\frac{128 u^2}{2145}\right)$\\
&$+\nu ^4 \left(\frac{64 u^{14}}{289575}+\frac{77632 u^{12}}{675675}+\frac{31321852 u^{10}}{6081075}+\frac{26430112 u^8}{675675}+\frac{7137596
   u^6}{155925}+\frac{1856 u^4}{10725}+\frac{224 u^2}{495}\right)$\\
&$+\nu ^3 \left(-\frac{1024 u^{14}}{4343625}-\frac{595168 u^{12}}{4343625}-\frac{64753184 u^{10}}{10135125}-\frac{493925008
   u^8}{10135125}-\frac{50372222 u^6}{921375}-\frac{631424 u^4}{675675}+\frac{43864 u^2}{45045}\right)$\\
&$+\nu ^2 \left(\frac{128 u^{14}}{1126125}+\frac{2188232
   u^{12}}{30405375}+\frac{109254434 u^{10}}{30405375}+\frac{293269616 u^8}{10135125}+\frac{296606029 u^6}{10135125}+\frac{42904 u^4}{61425}-\frac{1318 u^2}{15015}\right)$\\
&$+\nu 
   \left(-\frac{256 u^{14}}{10135125}-\frac{173984 u^{12}}{10135125}-\frac{28147738 u^{10}}{30405375}-\frac{81783508 u^8}{10135125}-\frac{51670481 u^6}{6756750}+\frac{57572
   u^4}{27027}-\frac{14117 u^2}{9009}\right)$\\
&$+\frac{64 u^{14}}{30405375}+\frac{15376 u^{12}}{10135125}+\frac{534004 u^{10}}{6081075}+\frac{550516 u^8}{675675}+\frac{473288
   u^6}{921375}-\frac{354176 u^4}{225225}+\frac{68608 u^2}{45045}$\\
$C_{01}^{(11)}$& $\nu ^5 \left(\frac{137728 u^{13}}{30405375}+\frac{1775744 u^{11}}{3378375}+\frac{25789472 u^9}{2764125}+\frac{6976096 u^7}{259875}+\frac{57472 u^5}{10125}+\frac{3584 u^3}{6435}\right)$\\
&$+\nu
   ^4 \left(-\frac{13184 u^{13}}{868725}-\frac{11185312 u^{11}}{6081075}-\frac{186957832 u^9}{6081075}-\frac{729688 u^7}{9009}-\frac{31128736 u^5}{2027025}+\frac{4736 u^3}{2925}\right)$\\
&$+\nu
   ^3 \left(\frac{75008 u^{13}}{4343625}+\frac{888736 u^{11}}{394875}+\frac{388650544 u^9}{10135125}+\frac{92189612 u^7}{921375}+\frac{230118632 u^5}{10135125}-\frac{94048
   u^3}{675675}\right)$\\
&$+\nu ^2 \left(-\frac{265472 u^{13}}{30405375}-\frac{12461968 u^{11}}{10135125}-\frac{681822074 u^9}{30405375}-\frac{9425048 u^7}{160875}-\frac{79484414
   u^5}{10135125}+\frac{1140808 u^3}{675675}\right)$\\
&$+\nu  \left(\frac{61184 u^{13}}{30405375}+\frac{9253144 u^{11}}{30405375}+\frac{402362 u^9}{66825}+\frac{154508516
   u^7}{10135125}-\frac{3490517 u^5}{3378375}+\frac{47756 u^3}{19305}\right)$\\
&$-\frac{5248 u^{13}}{30405375}-\frac{120992 u^{11}}{4343625}-\frac{18313028 u^9}{30405375}-\frac{259612
   u^7}{160875}-\frac{18896288 u^5}{10135125}-\frac{39168 u^3}{25025}$\\
$C_{11}^{(11)}$& $\nu ^5 \left(-\frac{256 u^{14}}{3378375}-\frac{2432 u^{12}}{78975}-\frac{12729616 u^{10}}{10135125}-\frac{27826048 u^8}{3378375}-\frac{56008352 u^6}{10135125}-\frac{140704
   u^4}{1447875}\right)$\\
&$+\nu ^4 \left(\frac{64 u^{14}}{289575}+\frac{653216 u^{12}}{6081075}+\frac{5208988 u^{10}}{1216215}+\frac{5789216 u^8}{225225}+\frac{4810444
   u^6}{289575}+\frac{4068104 u^4}{2027025}\right)$\\
&$+\nu ^3 \left(-\frac{1024 u^{14}}{4343625}-\frac{558176 u^{12}}{4343625}-\frac{14721344 u^{10}}{2764125}-\frac{108107836
   u^8}{3378375}-\frac{217083214 u^6}{10135125}+\frac{9786662 u^4}{10135125}\right)$\\
&$+\nu ^2 \left(\frac{128 u^{14}}{1126125}+\frac{158248 u^{12}}{2338875}+\frac{10173182
   u^{10}}{3378375}+\frac{1956004 u^8}{102375}+\frac{109689053 u^6}{10135125}+\frac{11885327 u^4}{20270250}\right)$\\
&$+\nu  \left(-\frac{256 u^{14}}{10135125}-\frac{44704
   u^{12}}{2764125}-\frac{23759774 u^{10}}{30405375}-\frac{2196791 u^8}{405405}-\frac{41173667 u^6}{20270250}+\frac{52197637 u^4}{13513500}\right)$\\
&$+\frac{64 u^{14}}{30405375}+\frac{14512
   u^{12}}{10135125}+\frac{323956 u^{10}}{4343625}+\frac{5554826 u^8}{10135125}-\frac{3029032 u^6}{10135125}-\frac{41835488 u^4}{10135125}-\frac{80896 u^2}{45045}$\\
\hline
\end{tabular}
\end{ruledtabular}
\end{table*}

\begin{table*}
\caption{\label{Cck1}  Energy spectrum at the tree level: list of coefficients at $p_\infty^{13}$}
\begin{ruledtabular}
\begin{tabular}{ll}
$C_{00}^{(13)}$& $\nu ^6 \left(\frac{64 u^{16}}{34459425}+\frac{504464 u^{14}}{310134825}+\frac{258852772 u^{12}}{1550674125}+\frac{1968391528 u^{10}}{516891375}+\frac{10672352 u^8}{530145}+\frac{1076398424
   u^6}{57432375}-\frac{256 u^4}{1989}-\frac{128 u^2}{3315}\right)$\\
&$+\nu ^5 \left(-\frac{256 u^{16}}{34459425}-\frac{3817024 u^{14}}{516891375}-\frac{34316752
   u^{12}}{46990125}-\frac{7730566576 u^{10}}{516891375}-\frac{1076657264 u^8}{15663375}-\frac{1925722856 u^6}{34459425}-\frac{117184 u^4}{546975}-\frac{10432 u^2}{36465}\right)$\\
&$+\nu ^4
   \left(\frac{128 u^{16}}{11486475}+\frac{6190432 u^{14}}{516891375}+\frac{17397608 u^{12}}{14768325}+\frac{11785646308 u^{10}}{516891375}+\frac{5469036236
   u^8}{57432375}+\frac{22751682859 u^6}{344594250}+\frac{466112 u^4}{883575}-\frac{70024 u^2}{109395}\right)$\\
&$+\nu ^3 \left(-\frac{128 u^{16}}{16081065}-\frac{98994976
   u^{14}}{10854718875}-\frac{9932792944 u^{12}}{10854718875}-\frac{62968238786 u^{10}}{3618239625}-\frac{1612610416 u^8}{24613875}-\frac{150734778329 u^6}{4824319500}-\frac{247292
   u^4}{11486475}-\frac{9032 u^2}{36465}\right)$\\
&$+\nu ^2 \left(\frac{64 u^{16}}{21928725}+\frac{12782384 u^{14}}{3618239625}+\frac{1326396028 u^{12}}{3618239625}+\frac{770741887
   u^{10}}{109643625}+\frac{3229592393 u^8}{141891750}+\frac{8648299241 u^6}{2412159750}-\frac{58025971 u^4}{80405325}+\frac{421 u^2}{1105}\right)$\\
&$+\nu  \left(-\frac{128
   u^{16}}{241215975}-\frac{2435936 u^{14}}{3618239625}-\frac{262803704 u^{12}}{3618239625}-\frac{312027523 u^{10}}{219287250}-\frac{93168371 u^8}{25525500}+\frac{12722399
   u^6}{8508500}-\frac{22531867 u^4}{17867850}+\frac{254671 u^2}{218790}\right)$\\
&$+\frac{64 u^{16}}{1688511825}+\frac{345872 u^{14}}{6907548375}+\frac{427332436
   u^{12}}{75983032125}+\frac{169207616 u^{10}}{1489863375}+\frac{1855672376 u^8}{8442559125}-\frac{141566624 u^6}{2814186375}+\frac{69702656 u^4}{80405325}-\frac{8192 u^2}{8415}$\\
$C_{01}^{(13)}$& $\nu ^6 \left(-\frac{3008 u^{15}}{19144125}-\frac{60215104 u^{13}}{1550674125}-\frac{2717924852 u^{11}}{1550674125}-\frac{1860689216 u^9}{103378275}-\frac{6147427744
   u^7}{172297125}-\frac{19114112 u^5}{3378375}-\frac{4096 u^3}{9945}\right)$\\
&$+\nu ^5 \left(\frac{1024 u^{15}}{1472625}+\frac{1175872 u^{13}}{6712875}+\frac{3764631488
   u^{11}}{516891375}+\frac{6754035152 u^9}{103378275}+\frac{491947616 u^7}{4417875}+\frac{102566144 u^5}{6891885}-\frac{681728 u^3}{546975}\right)$\\
&$+\nu ^4 \left(-\frac{3968
   u^{15}}{3614625}-\frac{49195456 u^{13}}{172297125}-\frac{657936472 u^{11}}{57432375}-\frac{48957226228 u^9}{516891375}-\frac{907986866 u^7}{6381375}-\frac{3326929192
   u^5}{172297125}-\frac{2080448 u^3}{11486475}\right)$\\
&$+\nu ^3 \left(\frac{25216 u^{15}}{30925125}+\frac{2396810656 u^{13}}{10854718875}+\frac{96308871608
   u^{11}}{10854718875}+\frac{14777430718 u^9}{212837625}+\frac{99361273261 u^7}{1206079875}+\frac{1412014756 u^5}{1206079875}-\frac{469712 u^3}{675675}\right)$\\
&$+\nu ^2 \left(-\frac{9536
   u^{15}}{30925125}-\frac{28681088 u^{13}}{328930875}-\frac{4330682768 u^{11}}{1206079875}-\frac{32346813659 u^9}{1206079875}-\frac{358236623 u^7}{19144125}+\frac{3192638782
   u^5}{1206079875}-\frac{7340684 u^3}{6185025}\right)$\\
&$+\nu  \left(\frac{128 u^{15}}{2226609}+\frac{61395392 u^{13}}{3618239625}+\frac{8072794 u^{11}}{11133045}+\frac{246104279
   u^9}{47297250}+\frac{125367967 u^7}{141891750}-\frac{5359481 u^5}{4060875}-\frac{27314906 u^3}{26801775}\right)$\\
&$-\frac{5056 u^{15}}{1206079875}-\frac{19578752
   u^{13}}{15196606425}-\frac{4356522284 u^{11}}{75983032125}-\frac{9911789716 u^9}{25327677375}+\frac{2364761488 u^7}{8442559125}+\frac{5667904384 u^5}{2814186375}+\frac{4462592
   u^3}{7309575}$\\
$C_{11}^{(13)}$& $\nu ^6 \left(\frac{64 u^{16}}{34459425}+\frac{2401936 u^{14}}{1550674125}+\frac{229917172 u^{12}}{1550674125}+\frac{4654752952 u^{10}}{1550674125}+\frac{2149423648
   u^8}{172297125}+\frac{1025812712 u^6}{172297125}+\frac{4558496 u^4}{57432375}\right)$\\
&$+\nu ^5 \left(-\frac{256 u^{16}}{34459425}-\frac{3639232 u^{14}}{516891375}-\frac{111329936
   u^{12}}{172297125}-\frac{6005611048 u^{10}}{516891375}-\frac{71767568 u^8}{1740375}-\frac{179222248 u^6}{10135125}-\frac{10062736 u^4}{6891885}\right)$\\
&$+\nu ^4 \left(\frac{128
   u^{16}}{11486475}+\frac{236384 u^{14}}{20675655}+\frac{59778872 u^{12}}{57432375}+\frac{1008691924 u^{10}}{57432375}+\frac{3198198742 u^8}{57432375}+\frac{661991653
   u^6}{31326750}-\frac{17273194 u^4}{13253625}\right)$\\
&$+\nu ^3 \left(-\frac{128 u^{16}}{16081065}-\frac{94612768 u^{14}}{10854718875}-\frac{798239168 u^{12}}{986792625}-\frac{144884918686
   u^{10}}{10854718875}-\frac{44521959739 u^8}{1206079875}-\frac{3801908797 u^6}{536035500}-\frac{784671554 u^4}{1206079875}\right)$\\
&$+\nu ^2 \left(\frac{64 u^{16}}{21928725}+\frac{12229808
   u^{14}}{3618239625}+\frac{391543588 u^{12}}{1206079875}+\frac{1499386517 u^{10}}{278326125}+\frac{4815838639 u^8}{402026625}-\frac{1776139 u^6}{877149}-\frac{6440454767
   u^4}{4824319500}\right)$\\
&$+\nu  \left(-\frac{128 u^{16}}{241215975}-\frac{777632 u^{14}}{1206079875}-\frac{233247416 u^{12}}{3618239625}-\frac{7893322153
   u^{10}}{7236479250}-\frac{3886356251 u^8}{2412159750}+\frac{2971235051 u^6}{2412159750}-\frac{77835629 u^4}{21021000}\right)$\\
&$+\frac{64 u^{16}}{1688511825}+\frac{3646768
   u^{14}}{75983032125}+\frac{380170564 u^{12}}{75983032125}+\frac{30003968 u^{10}}{343814625}+\frac{596951588 u^8}{8442559125}+\frac{2147912512 u^6}{8442559125}+\frac{8921101184
   u^4}{2814186375}+\frac{950272 u^2}{765765}$\\
\hline
\end{tabular}
\end{ruledtabular}
\end{table*}

\begin{table*}
\caption{\label{Cck2}  Energy spectrum at the tree level: list of coefficients at $p_\infty^{15}$}
\begin{ruledtabular}
\begin{tabular}{ll}
$C_{00}^{(15)}$& $\nu ^7 \left(-\frac{128 u^{18}}{3749811975}-\frac{3707488 u^{16}}{68746552875}-\frac{2218977256 u^{14}}{206239658625}-\frac{6588802768 u^{12}}{12131744625}-\frac{2564384
   u^{10}}{326781}-\frac{19622464112 u^8}{654729075}-\frac{45752355632 u^6}{2083228875}+\frac{512 u^4}{4845}+\frac{128 u^2}{4845}\right)$\\
&$+\nu ^6 \left(\frac{32
   u^{18}}{178562475}+\frac{1835192 u^{16}}{5892561675}+\frac{1753481398 u^{14}}{29462808375}+\frac{3164103476 u^{12}}{1178512335}+\frac{21699664132 u^{10}}{654729075}+\frac{13822782412
   u^8}{130945815}+\frac{211184887336 u^6}{3273645375}+\frac{4672 u^4}{20995}+\frac{12256 u^2}{62985}\right)$\\
&$+\nu ^5 \left(-\frac{64 u^{18}}{178562475}-\frac{498544
   u^{16}}{755456625}-\frac{1207875308 u^{14}}{9820936125}-\frac{16769640856 u^{12}}{3273645375}-\frac{549520409638 u^{10}}{9820936125}-\frac{494099212462
   u^8}{3273645375}-\frac{47017844138 u^6}{654729075}\right.$\\
&$\left. -\frac{4127456 u^4}{14549535}+\frac{317024 u^2}{692835}\right)$\\
&$+\nu ^4 \left(\frac{88 u^{18}}{249987465}+\frac{400094
   u^{16}}{587577375}+\frac{946677203 u^{14}}{7499623950}+\frac{158621917837 u^{12}}{31729178250}+\frac{51797928508 u^{10}}{1057639275}+\frac{4796117369987
   u^8}{45831035250}+\frac{2361879986381 u^6}{91662070500}\right.$\\
&$\left.-\frac{11512532 u^4}{72747675}+\frac{51056 u^2}{159885}\right)$\\
&$+\nu ^3 \left(-\frac{704 u^{18}}{3749811975}-\frac{222848
   u^{16}}{587577375}-\frac{2924599064 u^{14}}{41247931725}-\frac{563201610476 u^{12}}{206239658625}-\frac{126041932736 u^{10}}{5288196375}-\frac{56055926159
   u^8}{1666583100}+\frac{1209369023 u^6}{564074280}\right.$\\
&$\left.+\frac{1721084 u^4}{7309575}-\frac{382061 u^2}{14549535}\right)$\\
&$+\nu ^2 \left(\frac{16 u^{18}}{288447075}+\frac{615532
   u^{16}}{5288196375}+\frac{10347397 u^{14}}{467663625}+\frac{3858599902 u^{12}}{4583103525}+\frac{62871949022 u^{10}}{9820936125}+\frac{629140882537 u^8}{183324141000}-\frac{26383399589
   u^6}{16665831000}\right.$\\
&$\left. +\frac{563730641 u^4}{1018467450}-\frac{4580131 u^2}{11639628}\right)$\\
&$+\nu  \left(-\frac{32 u^{18}}{3749811975}-\frac{2969864 u^{16}}{160408623375}-\frac{5189281282
   u^{14}}{1443677610375}-\frac{197207177786 u^{12}}{1443677610375}-\frac{421878827303 u^{10}}{481225870125}+\frac{173300811923 u^8}{513307594800}-\frac{135453272281
   u^6}{285170886000}\right.$\\
&$\left. +\frac{4767513203 u^4}{6110804700}-\frac{4791731 u^2}{5542680}\right)$\\
&$+\frac{2 u^{18}}{3749811975}+\frac{4201 u^{16}}{3525464250}+\frac{210322499
   u^{14}}{888416991000}+\frac{104453063983 u^{12}}{11549420883000}+\frac{22426121147 u^{10}}{481225870125}-\frac{11718735568 u^8}{160408623375}-\frac{191321024
   u^6}{2138781645}$\\
&$-\frac{159711232 u^4}{305540235}+\frac{9715712 u^2}{14549535}$\\
$C_{01}^{(15)}$& $\nu ^7 \left(\frac{4096 u^{17}}{1057639275}+\frac{369548768 u^{15}}{206239658625}+\frac{237085136 u^{13}}{1402990875}+\frac{896820715528 u^{11}}{206239658625}+\frac{99184221152
   u^9}{3273645375}+\frac{114706756496 u^7}{2546168625}+\frac{11738294272 u^5}{2083228875}+\frac{512 u^3}{1615}\right)$\\
&$+\nu ^6 \left(-\frac{43136 u^{17}}{1964187225}-\frac{33511976
   u^{15}}{3273645375}-\frac{458611772 u^{13}}{516891375}-\frac{583526117296 u^{11}}{29462808375}-\frac{22178126216 u^9}{192567375}-\frac{7315726084 u^7}{51962625}-\frac{46271617856
   u^5}{3273645375}+\frac{63616 u^3}{62985}\right)$\\
&$+\nu ^5 \left(\frac{17984 u^{17}}{392837445}+\frac{212050624 u^{15}}{9820936125}+\frac{5788502116 u^{13}}{3273645375}+\frac{348669991336
   u^{11}}{9820936125}+\frac{31674373708 u^9}{178562475}+\frac{14711210498 u^7}{83939625}+\frac{10198526896 u^5}{654729075}+\frac{286976 u^3}{765765}\right)$\\
&$+\nu ^4 \left(-\frac{127744
   u^{17}}{2749862115}-\frac{184429646 u^{15}}{8249586345}-\frac{122209564049 u^{13}}{68746552875}-\frac{1359302508191 u^{11}}{41247931725}-\frac{1907717304034
   u^9}{13749310575}-\frac{1344591272201 u^7}{15277011750}+\frac{83444865776 u^5}{22915517625}+\frac{3206144 u^3}{8083075}\right)$\\
&$+\nu ^3 \left(\frac{95392
   u^{17}}{3749811975}+\frac{863295896 u^{15}}{68746552875}+\frac{2721694994 u^{13}}{2749862115}+\frac{3549342746284 u^{11}}{206239658625}+\frac{3890831426813
   u^9}{68746552875}+\frac{30594854203 u^7}{3666482820}-\frac{6981545119 u^5}{4583103525}+\frac{1010206412 u^3}{1527701175}\right)$\\
&$+\nu ^2 \left(-\frac{316864
   u^{17}}{41247931725}-\frac{53590708 u^{15}}{13749310575}-\frac{21226606318 u^{13}}{68746552875}-\frac{49916562848 u^{11}}{9820936125}-\frac{152235850381
   u^9}{13749310575}+\frac{427725937 u^7}{94594500}-\frac{79143856073 u^5}{91662070500}+\frac{19272289 u^3}{26801775}\right)$\\
&$+\nu  \left(\frac{347552 u^{17}}{288735522075}+\frac{181251088
   u^{15}}{288735522075}+\frac{24255713402 u^{13}}{481225870125}+\frac{1139382716623 u^{11}}{1443677610375}+\frac{38278740257 u^9}{56614808250}-\frac{1405451219441
   u^7}{1283268987000}+\frac{772983196459 u^5}{427756329000}+\frac{988809113 u^3}{3055402350}\right)$\\
&$-\frac{3152 u^{17}}{41247931725}-\frac{39381071 u^{15}}{962451740250}-\frac{6441774767
   u^{13}}{1924903480500}-\frac{289166728801 u^{11}}{5774710441500}+\frac{12989963176 u^9}{481225870125}-\frac{1540170848 u^7}{17823180375}-\frac{768792320 u^5}{427756329}-\frac{60514304
   u^3}{305540235}$\\
$C_{11}^{(15)}$& $\nu ^7 \left(-\frac{128 u^{18}}{3749811975}-\frac{10726624 u^{16}}{206239658625}-\frac{679823288 u^{14}}{68746552875}-\frac{8687498828 u^{12}}{18749059875}-\frac{1214354997344
   u^{10}}{206239658625}-\frac{132946702288 u^8}{7638505875}-\frac{144905627792 u^6}{22915517625}-\frac{135895328 u^4}{2083228875}\right)$\\
&$+\nu ^6 \left(\frac{32
   u^{18}}{178562475}+\frac{136232 u^{16}}{453273975}+\frac{2318942 u^{14}}{42514875}+\frac{66802141543 u^{12}}{29462808375}+\frac{143111237956 u^{10}}{5892561675}+\frac{21287861908
   u^8}{363738375}+\frac{59277268016 u^6}{3273645375}+\frac{278254888 u^4}{251818875}\right)$\\
&$+\nu ^5 \left(-\frac{64 u^{18}}{178562475}-\frac{568912 u^{16}}{892812375}-\frac{1104924572
   u^{14}}{9820936125}-\frac{6022154344 u^{12}}{1402990875}-\frac{35870243302 u^{10}}{892812375}-\frac{86916199742 u^8}{1091215125}-\frac{4827060146 u^6}{251818875}+\frac{908071624
   u^4}{654729075}\right)$\\
&$+\nu ^4 \left(\frac{88 u^{18}}{249987465}+\frac{3478946 u^{16}}{5288196375}+\frac{15863287103 u^{14}}{137493105750}+\frac{62480972293
   u^{12}}{14999247900}+\frac{1419279648316 u^{10}}{41247931725}+\frac{770927561381 u^8}{15277011750}+\frac{12522323951 u^6}{5391886500}+\frac{20023132436 u^4}{22915517625}\right)$\\
&$+\nu ^3
   \left(-\frac{704 u^{18}}{3749811975}-\frac{4447984 u^{16}}{12131744625}-\frac{2673044576 u^{14}}{41247931725}-\frac{467532695719 u^{12}}{206239658625}-\frac{3355015724288
   u^{10}}{206239658625}-\frac{1157643253801 u^8}{91662070500}+\frac{2321111669 u^6}{643242600}+\frac{15396124277 u^4}{18332414100}\right)$\\
&$+\nu ^2 \left(\frac{16
   u^{18}}{288447075}+\frac{3317044 u^{16}}{29462808375}+\frac{198699481 u^{14}}{9820936125}+\frac{1742956877 u^{12}}{2499874650}+\frac{287312103193 u^{10}}{68746552875}-\frac{41029853
   u^8}{102702600}+\frac{9512946059 u^6}{16665831000}+\frac{550220440243 u^4}{366648282000}\right)$\\
&$+\nu  \left(-\frac{32 u^{18}}{3749811975}-\frac{25866056
   u^{16}}{1443677610375}-\frac{45227386 u^{14}}{13749310575}-\frac{1792274956 u^{12}}{15864589125}-\frac{1540925192431 u^{10}}{2887355220750}+\frac{190584029899
   u^8}{366648282000}-\frac{130747533493 u^6}{197425998000}+\frac{5543686631671 u^4}{1711025316000}\right)$\\
&$+\frac{2 u^{18}}{3749811975}+\frac{67981 u^{16}}{58925616750}+\frac{2501959
   u^{14}}{11537883000}+\frac{172654434721 u^{12}}{23098841766000}+\frac{163169071 u^{10}}{6532477875}-\frac{239821768 u^8}{4113041625}-\frac{4658741888 u^6}{17823180375}-\frac{5343672064
   u^4}{2138781645}-\frac{2670592 u^2}{2909907}$\\
\hline
\end{tabular}
\end{ruledtabular}
\end{table*}

Concerning the 1-loop energy spectrum (in the cm-frame), we have the structure:
\bea
{\mathcal F}^{\rm 1 loop}(u)&=&\frac{1}{p_\infty}\underbrace{{\mathcal F}^{\rm 1 loop}_{-1}(u)}_{\rm Newtonian}+{\mathcal F}^{\rm 1 loop}_{0}(u)+{\mathcal F}^{\rm 1 loop}_{1}(u)p_\infty\nonumber\\
&+&{\mathcal F}^{\rm 1 loop}_{2}(u)p_\infty^2 +\ldots + \underbrace{{\mathcal F}^{\rm 1 loop}_{14}(u)}_{\rm 7.5PN}p_\infty^{14} \nonumber\\
&+&O(p_\infty^{15}).
\eea
Each coefficient ${\mathcal F}^{\rm 1 loop}_{k}(u)$ is bilinear in $K_0(u)$,  $K_1(u)$, and $e^{-u}$ (which comes from Bessel K functions with half-integer indices),
namely:
\bea
{\mathcal F}^{\rm 1 loop}_{k}(u)&=& \frac{G^3  M^5  \nu^2}{b^4} \left[ A_{00}^{(k)} K_0(u)^2\right.\nonumber\\
&+&A_{01}^{(k)} K_0(u)K_1(u)+ A_{11}^{(k)} K_1(u)^2+\nonumber\\
&+& \left.e^{-u}\left(A_{0{\rm exp}}^{(k)} K_0(u)+A_{1{\rm exp}}^{(k)} K_1(u)\right) \right]
\eea
For example, at leading order, $k=-1$, the $e^{-u}$ terms are absent and
\bea
A_{00}^{(-1)}&=& \frac{32 u^5}{5}+\frac{32 u^3}{15}\nonumber\\
A_{01}^{(-1)}&=& \frac{96}{5} u^4\nonumber\\
A_{11}^{(-1)}&=& \frac{32 u^5}{5}+\frac{32 u^3}{5}\nonumber\\
A_{0{\rm exp}}^{(-1)}&=& 0 \nonumber\\
A_{1{\rm exp}}^{(-1)}&=& 0\,.
\eea
For $k=0$ all coefficients vanish, while for $k=1$ we find
\bea
A_{00}^{(1)}&=& \nu  \left(-\frac{64 u^7}{21}+\frac{16 u^5}{5}+\frac{16 u^3}{7}\right)\nonumber\\
&+& \frac{16 u^7}{21}-\frac{992 u^5}{105}-\frac{1136 u^3}{105}\nonumber\\
A_{01}^{(1)}&=& \nu  \left(\frac{128 u^6}{35}+\frac{304 u^4}{21}\right)\nonumber\\
&-&\frac{736 u^6}{105}-\frac{144 u^4}{7}\nonumber\\
A_{11}^{(1)}&=& \nu  \left(-\frac{64 u^7}{21}+\frac{368 u^5}{105}+\frac{16 u^3}{5}\right)\nonumber\\
&+& \frac{16 u^7}{21}-\frac{88 u^5}{7}+\frac{464 u^3}{35}\nonumber\\
A_{0{\rm exp}}^{(1)}&=& \frac{96 u^4}{5}+\frac{144 u^3}{5}+\frac{144 u^2}{5}\nonumber\\
A_{1{\rm exp}}^{(1)}&=& \frac{96 u^4}{5}+\frac{192 u^3}{5}+\frac{96 u^2}{5}+\frac{96 u}{5}\,.
\eea
Table \ref{Aabk} below contains the corresponding coefficients at high PN orders.

\begin{table*}
\caption{\label{Aabk}  Energy spectrum at 1-loop: list of coefficients from $p_\infty^2$ up to $p_\infty^6$.}
\begin{ruledtabular}
\begin{tabular}{ll}
$A_{00}^{(2)}$ &$ \frac{64 u^5}{5}+\frac{64 u^3}{15}$\\
$A_{01}^{(2)}$ &$ \frac{192}{5} u^4 $\\
$A_{11}^{(2)}$ &$\frac{64 u^5}{5}+\frac{64 u^3}{5}$\\
$A_{0{\rm exp}}^{(2)}$&$0$\\
$A_{1{\rm exp}}^{(2)}$&$0$\\
\hline
$A_{00}^{(3)}$ &$\nu ^2 \left(\frac{32 u^9}{63}+\frac{4136 u^7}{945}-\frac{1012 u^5}{945}-\frac{4 u^3}{63}\right)+\nu  \left(-\frac{64 u^9}{189}-\frac{232 u^7}{63}-\frac{596 u^5}{189}-\frac{548
   u^3}{45}\right)+\frac{32 u^9}{567}+\frac{4352 u^7}{2835}-\frac{1388 u^5}{105}+\frac{4252 u^3}{315} $\\
$A_{01}^{(3)}$ &$ \nu ^2 \left(-\frac{928 u^8}{189}-\frac{1088 u^6}{315}-\frac{2012 u^4}{945}\right)+\nu  \left(\frac{128 u^8}{21}+\frac{2896 u^6}{945}-\frac{26324 u^4}{945}\right)-\frac{32
   u^8}{27}+\frac{23888 u^6}{2835}-\frac{2228 u^4}{63}$\\
$A_{11}^{(3)}$ &$ \nu ^2 \left(\frac{32 u^9}{63}+\frac{2056 u^7}{945}-\frac{172 u^5}{315}-\frac{4 u^3}{5}\right)+\nu  \left(-\frac{64 u^9}{189}-\frac{152 u^7}{189}-\frac{2920 u^5}{189}+\frac{204
   u^3}{35}\right)+\frac{32 u^9}{567}+\frac{2752 u^7}{2835}+\frac{23456 u^5}{2835}-\frac{7948 u^3}{315}$\\
$A_{0{\rm exp}}^{(3)}$&$\nu  \left(-\frac{64 u^6}{7}+\frac{64 u^5}{7}+\frac{152 u^4}{35}+\frac{288 u^3}{35}+\frac{288 u^2}{35}\right)+\frac{16 u^6}{7}-\frac{572 u^5}{35}-\frac{208 u^4}{35}+\frac{44
   u^3}{35}+\frac{44 u^2}{35} $\\
$A_{1{\rm exp}}^{(3)}$&$\nu  \left(-\frac{64 u^6}{7}+\frac{32 u^5}{7}+\frac{192 u^4}{35}+\frac{32 u^3}{5}\right)+\frac{16 u^6}{7}-\frac{76 u^5}{5}-\frac{464 u^4}{35}+\frac{1128 u^3}{35}+\frac{1896
   u^2}{35}+\frac{1896 u}{35} $\\
\hline
$A_{00}^{(4)}$ &$\nu  \left(-\frac{128 u^7}{21}+\frac{32 u^5}{5}+\frac{32 u^3}{7}\right)+\frac{32 u^7}{21}+\frac{32 u^5}{105}-\frac{320 u^3}{21}$\\
$A_{01}^{(4)}$ &$\nu  \left(\frac{256 u^6}{35}+\frac{608 u^4}{21}\right)-\frac{1472 u^6}{105}+\frac{576 u^4}{35}$\\
$A_{11}^{(4)}$ &$\nu  \left(-\frac{128 u^7}{21}+\frac{736 u^5}{105}+\frac{32 u^3}{5}\right)+\frac{32 u^7}{21}-\frac{208 u^5}{35}+\frac{320 u^3}{7}$\\
$A_{0{\rm exp}}^{(4)}$&$0$\\
$A_{1{\rm exp}}^{(4)}$&$0$\\
\hline
$A_{00}^{(5)}$ &$\nu ^3 \left(-\frac{64 u^{11}}{1485}-\frac{73312 u^9}{31185}-\frac{179044 u^7}{31185}+\frac{118 u^5}{231}-\frac{26 u^3}{693}\right)+\nu ^2 \left(\frac{16 u^{11}}{297}+\frac{42172
   u^9}{10395}+\frac{115078 u^7}{10395}+\frac{5738 u^5}{2079}+\frac{1196 u^3}{3465}\right)$\\&$+\nu  \left(-\frac{32 u^{11}}{1485}-\frac{67016 u^9}{31185}+\frac{626 u^7}{1155}-\frac{124016
   u^5}{10395}+\frac{12286 u^3}{693}\right)+\frac{4 u^{11}}{1485}+\frac{9697 u^9}{31185}-\frac{20896 u^7}{10395}+\frac{5744 u^5}{495}+\frac{5570 u^3}{693} $\\
$A_{01}^{(5)}$ &$ \nu ^3 \left(\frac{3328 u^{10}}{3465}+\frac{274528 u^8}{31185}+\frac{78224 u^6}{31185}+\frac{2558 u^4}{3465}\right)+\nu ^2 \left(-\frac{128 u^{10}}{77}-\frac{100 u^8}{7}-\frac{9844
   u^6}{3465}+\frac{26084 u^4}{10395}\right)$\\
&$+\nu  \left(\frac{4640 u^{10}}{6237}+\frac{164896 u^8}{31185}+\frac{24572 u^6}{10395}-\frac{94942 u^4}{10395}\right)-\frac{208
   u^{10}}{2079}-\frac{30803 u^8}{31185}+\frac{40924 u^6}{10395}-\frac{55094 u^4}{3465}$\\
$A_{11}^{(5)}$ &$\nu ^3 \left(-\frac{64 u^{11}}{1485}-\frac{59008 u^9}{31185}-\frac{68212 u^7}{31185}+\frac{8678 u^5}{31185}+\frac{2 u^3}{5}\right)+\nu ^2 \left(\frac{16 u^{11}}{297}+\frac{33812
   u^9}{10395}+\frac{6206 u^7}{1155}+\frac{1061 u^5}{495}-\frac{44 u^3}{35}\right)$\\
&$+\nu  \left(-\frac{32 u^{11}}{1485}-\frac{18584 u^9}{10395}+\frac{74038 u^7}{31185}+\frac{136541
   u^5}{10395}-\frac{874 u^3}{63}\right)+\frac{4 u^{11}}{1485}+\frac{8179 u^9}{31185}-\frac{74348 u^7}{31185}-\frac{3161 u^5}{231}-\frac{185978 u^3}{3465} $\\
$A_{0{\rm exp}}^{(5)}$&$\nu ^2 \left(\frac{32 u^8}{21}-\frac{544 u^7}{63}+\frac{1100 u^6}{63}-\frac{3254 u^5}{315}-\frac{440 u^4}{63}-\frac{352 u^3}{105}-\frac{352 u^2}{105}\right)$\\
&$+\nu  \left(-\frac{64
   u^8}{63}+\frac{104 u^7}{9}-\frac{3382 u^6}{105}+\frac{6904 u^5}{315}+\frac{1426 u^4}{63}-\frac{232 u^3}{15}-\frac{232 u^2}{15}\right)$\\
&$+\frac{32 u^8}{189}-\frac{436
   u^7}{189}+\frac{20617 u^6}{1575}-\frac{7547 u^5}{1575}-\frac{27457 u^4}{1575}-\frac{17812 u^3}{1575}-\frac{548 u^2}{35} $\\
$A_{1{\rm exp}}^{(5)}$&$ \nu ^2 \left(\frac{32 u^8}{21}-\frac{496 u^7}{63}+\frac{96 u^6}{7}-\frac{64 u^5}{15}-\frac{296 u^4}{315}-\frac{704 u^3}{315}\right)+\nu  \left(-\frac{64 u^8}{63}+\frac{232
   u^7}{21}-\frac{8446 u^6}{315}+\frac{3076 u^5}{315}-\frac{5408 u^4}{315}-\frac{6008 u^3}{315}\right)$\\
&$+\frac{32 u^8}{189}-\frac{20 u^7}{9}+\frac{56701 u^6}{4725}+\frac{1499
   u^5}{1575}+\frac{10874 u^4}{1575}+\frac{694 u^3}{105}+\frac{392 u^2}{15}+\frac{392 u}{15} $\\
\hline
$A_{00}^{(6)}$ &$ \nu ^2 \left(\frac{64 u^9}{63}+\frac{8272 u^7}{945}-\frac{2024 u^5}{945}-\frac{8 u^3}{63}\right)+\nu  \left(-\frac{128 u^9}{189}-\frac{1040 u^7}{63}+\frac{3112 u^5}{945}-\frac{5512
   u^3}{315}\right)+\frac{64 u^9}{567}+\frac{15184 u^7}{2835}-\frac{1216 u^5}{105}+\frac{2816 u^3}{315}$\\
$A_{01}^{(6)}$ &$\nu ^2 \left(-\frac{1856 u^8}{189}-\frac{2176 u^6}{315}-\frac{4024 u^4}{945}\right)+\nu  \left(\frac{256 u^8}{21}+\frac{3232 u^6}{189}-\frac{11608 u^4}{945}\right)-\frac{64
   u^8}{27}-\frac{2368 u^6}{567}-\frac{128 u^4}{45} $\\
$A_{11}^{(6)}$ &$\nu ^2 \left(\frac{64 u^9}{63}+\frac{4112 u^7}{945}-\frac{344 u^5}{315}-\frac{8 u^3}{5}\right)+\nu  \left(-\frac{128 u^9}{189}-\frac{2032 u^7}{189}-\frac{2752 u^5}{135}+\frac{744
   u^3}{35}\right)+\frac{64 u^9}{567}+\frac{1712 u^7}{405}+\frac{62464 u^5}{2835}+\frac{2048 u^3}{63} $\\
$A_{0{\rm exp}}^{(6)}$&$0 $\\
$A_{1{\rm exp}}^{(6)}$&$0$\\
\hline
\end{tabular}
\end{ruledtabular}
\end{table*}

\begin{table*}
\caption{\label{Aabk}  Energy spectrum at 1-loop: list of coefficients at $p_\infty^7$.}
\begin{ruledtabular}
\begin{tabular}{ll}
$A_{00}^{(7)}$ &$\nu ^4 \left(\frac{128 u^{13}}{57915}+\frac{3488 u^{11}}{9009}+\frac{37892 u^9}{7371}+\frac{922717 u^7}{135135}-\frac{52597 u^5}{180180}+\frac{439 u^3}{12012}\right)$\\
&$+\nu ^3
   \left(-\frac{256 u^{13}}{57915}-\frac{6056 u^{11}}{6237}-\frac{4739698 u^9}{405405}-\frac{456674 u^7}{31185}-\frac{140879 u^5}{90090}+\frac{1819 u^3}{8190}\right)$\\
&$+\nu ^2 \left(\frac{896
   u^{13}}{289575}+\frac{321064 u^{11}}{405405}+\frac{3147766 u^9}{405405}-\frac{315736 u^7}{225225}-\frac{24907 u^5}{77220}-\frac{95813 u^3}{180180}\right)$\\
&$+\nu  \left(-\frac{256
   u^{13}}{289575}-\frac{6766 u^{11}}{27027}-\frac{87027 u^9}{50050}+\frac{10648567 u^7}{675675}+\frac{1315421 u^5}{245700}+\frac{116243 u^3}{36036}\right)$\\
&$+\frac{128
   u^{13}}{1447875}+\frac{91582 u^{11}}{3378375}+\frac{329921 u^9}{2252250}-\frac{14789821 u^7}{3378375}+\frac{925757 u^5}{128700}-\frac{3362911 u^3}{180180}  $\\
$A_{01}^{(7)}$ &$ \nu ^4 \left(-\frac{128 u^{12}}{1485}-\frac{1360256 u^{10}}{405405}-\frac{5030548 u^8}{405405}-\frac{267976 u^6}{135135}-\frac{63839 u^4}{180180}\right)$\\
&$+\nu ^3 \left(\frac{12224
   u^{12}}{57915}+\frac{1093712 u^{10}}{135135}+\frac{1151606 u^8}{45045}+\frac{191254 u^6}{57915}-\frac{33647 u^4}{90090}\right)$\\
&$+\nu ^2 \left(-\frac{46528 u^{12}}{289575}-\frac{12873152
   u^{10}}{2027025}-\frac{17785582 u^8}{2027025}-\frac{238372 u^6}{675675}+\frac{2474809 u^4}{540540}\right)$\\
&$+\nu  \left(\frac{14096 u^{12}}{289575}+\frac{1402468
   u^{10}}{675675}-\frac{1722701 u^8}{270270}-\frac{3269534 u^6}{135135}+\frac{584351 u^4}{386100}\right)$\\
&$-\frac{1472 u^{12}}{289575}-\frac{36808 u^{10}}{160875}+\frac{2305753
   u^8}{965250}+\frac{32531606 u^6}{3378375}+\frac{8223791 u^4}{300300} $\\
$A_{11}^{(7)}$ &$  \nu ^4 \left(\frac{128 u^{13}}{57915}+\frac{139936 u^{11}}{405405}+\frac{293996 u^9}{81081}+\frac{125117 u^7}{57915}-\frac{27 u^5}{140}-\frac{u^3}{4}\right)$\\
&$+\nu ^3 \left(-\frac{256
   u^{13}}{57915}-\frac{351752 u^{11}}{405405}-\frac{155486 u^9}{19305}-\frac{2273734 u^7}{405405}-\frac{221084 u^5}{405405}+\frac{37 u^3}{70}\right)$\\
&$+\nu ^2 \left(\frac{896
   u^{13}}{289575}+\frac{1445608 u^{11}}{2027025}+\frac{9987482 u^9}{2027025}-\frac{2905444 u^7}{2027025}-\frac{11549233 u^5}{2702700}+\frac{4703 u^3}{1260}\right)$\\
&$+\nu  \left(-\frac{256
   u^{13}}{289575}-\frac{91802 u^{11}}{405405}-\frac{364213 u^9}{450450}+\frac{87727 u^7}{9009}+\frac{2606537 u^5}{540540}-\frac{106741 u^3}{4620}\right)$\\
&$+\frac{128
   u^{13}}{1447875}+\frac{249434 u^{11}}{10135125}+\frac{4553 u^9}{103950}-\frac{592897 u^7}{225225}-\frac{72093121 u^5}{13513500}-\frac{169133 u^3}{180180}$\\
$A_{0{\rm exp}}^{(7)}$&$\nu ^3 \left(-\frac{64 u^{10}}{495}+\frac{1856 u^9}{1155}-\frac{9764 u^8}{1155}+\frac{14948 u^7}{693}-\frac{8147 u^6}{315}+\frac{36191 u^5}{3465}+\frac{5288 u^4}{693}+\frac{416
   u^3}{231}+\frac{416 u^2}{231}\right)$\\
&$+\nu ^2 \left(\frac{16 u^{10}}{99}-\frac{1964 u^9}{693}+\frac{185881 u^8}{10395}-\frac{17621 u^7}{330}+\frac{98635 u^6}{1386}-\frac{1446
   u^5}{77}-\frac{27908 u^4}{1155}+\frac{496 u^3}{77}+\frac{496 u^2}{77}\right)$\\
&$+\nu  \left(-\frac{32 u^{10}}{495}+\frac{95248 u^9}{72765}-\frac{783626 u^8}{72765}+\frac{627967
   u^7}{17325}-\frac{6584582 u^6}{121275}+\frac{543539 u^5}{40425}+\frac{326852 u^4}{11025}-\frac{40496 u^3}{40425}-\frac{1616 u^2}{385}\right)$\\
&$+\frac{4 u^{10}}{495}-\frac{4427
   u^9}{24255}+\frac{23393 u^8}{13860}-\frac{2991739 u^7}{363825}+\frac{1175378 u^6}{121275}-\frac{49547 u^5}{24255}-\frac{452554 u^4}{121275}-\frac{553144 u^3}{121275}+\frac{5416
   u^2}{1155}  $\\
$A_{1{\rm exp}}^{(7)}$&$ \nu ^3 \left(-\frac{64 u^{10}}{495}+\frac{5344 u^9}{3465}-\frac{2964 u^8}{385}+\frac{62014 u^7}{3465}-\frac{61364 u^6}{3465}+\frac{3596 u^5}{1155}-\frac{4 u^4}{495}+\frac{736
   u^3}{693}\right)$\\
&$+\nu ^2 \left(\frac{16 u^{10}}{99}-\frac{212 u^9}{77}+\frac{171781 u^8}{10395}-\frac{135011 u^7}{2970}+\frac{115937 u^6}{2310}-\frac{28334 u^5}{3465}+\frac{4324
   u^4}{385}+\frac{10456 u^3}{1155}\right)$\\
&$+\nu  \left(-\frac{32 u^{10}}{495}+\frac{92896 u^9}{72765}-\frac{81974 u^8}{8085}+\frac{11398012 u^7}{363825}-\frac{1928317
   u^6}{48510}+\frac{1010336 u^5}{40425}+\frac{478528 u^4}{121275}-\frac{5728 u^3}{693}\right)$\\
&$+\frac{4 u^{10}}{495}-\frac{481 u^9}{2695}+\frac{155191 u^8}{97020}-\frac{984911
   u^7}{132300}+\frac{639167 u^6}{103950}-\frac{93001 u^5}{11025}-\frac{54596 u^4}{121275}-\frac{7412 u^3}{1155}-\frac{1328 u^2}{165}-\frac{1328 u}{165} $\\
\end{tabular}
\end{ruledtabular}
\end{table*}

\begin{table*}
\caption{\label{Aabk}  Energy spectrum at 1-loop: list of coefficients at $p_\infty^8$.}
\begin{ruledtabular}
\begin{tabular}{ll}
$A_{00}^{(8)}$ &$ \nu ^3 \left(-\frac{128 u^{11}}{1485}-\frac{146624 u^9}{31185}-\frac{358088 u^7}{31185}+\frac{236 u^5}{231}-\frac{52 u^3}{693}\right)
+\nu ^2 \left(\frac{32 u^{11}}{297}+\frac{14312
   u^9}{1485}+\frac{366644 u^7}{10395}+\frac{23984 u^5}{10395}+\frac{1732 u^3}{3465}\right)$\\
&$+\nu  \left(-\frac{64 u^{11}}{1485}-\frac{165712 u^9}{31185}-\frac{11756 u^7}{385}-\frac{24368
   u^5}{2079}+\frac{49732 u^3}{3465}\right)+\frac{8 u^{11}}{1485}+\frac{24674 u^9}{31185}+\frac{11908 u^7}{2079}+\frac{7552 u^5}{3465}+\frac{38144 u^3}{3465} $\\
$A_{01}^{(8)}$ &$  \nu ^3 \left(\frac{6656 u^{10}}{3465}+\frac{549056 u^8}{31185}+\frac{156448 u^6}{31185}+\frac{5116 u^4}{3465}\right)+\nu ^2 \left(-\frac{256 u^{10}}{77}-\frac{2728 u^8}{63}-\frac{55592
   u^6}{3465}-\frac{14228 u^4}{10395}\right)$\\
&$+\nu  \left(\frac{9280 u^{10}}{6237}+\frac{128576 u^8}{4455}+\frac{80264 u^6}{2079}-\frac{42836 u^4}{10395}\right)-\frac{416
   u^{10}}{2079}-\frac{172486 u^8}{31185}-\frac{49072 u^6}{3465}-\frac{20480 u^4}{693}$\\
$A_{11}^{(8)}$ &$ \nu ^3 \left(-\frac{128 u^{11}}{1485}-\frac{118016 u^9}{31185}-\frac{136424 u^7}{31185}+\frac{17356 u^5}{31185}+\frac{4 u^3}{5}\right)+\nu ^2 \left(\frac{32 u^{11}}{297}+\frac{83464
   u^9}{10395}+\frac{59852 u^7}{3465}+\frac{9178 u^5}{3465}-\frac{172 u^3}{35}\right)$\\
&$+\nu  \left(-\frac{64 u^{11}}{1485}-\frac{47728 u^9}{10395}-\frac{568684 u^7}{31185}+\frac{5314
   u^5}{1485}+\frac{3572 u^3}{315}\right)+\frac{8 u^{11}}{1485}+\frac{21638 u^9}{31185}+\frac{20500 u^7}{6237}-\frac{52016 u^5}{10395}-\frac{38656 u^3}{3465} $\\
$A_{0{\rm exp}}^{(8)}$&$ 0 $\\
$A_{1{\rm exp}}^{(8)}$&$ 0 $\\
\end{tabular}
\end{ruledtabular}
\end{table*}

\begin{table*}
\caption{\label{Aabk}  Energy spectrum at 1-loop: list of coefficients at $p_\infty^9$.}
\begin{ruledtabular}
\begin{tabular}{ll}
$A_{00}^{(9)}$ &$\nu ^5 \left(-\frac{256 u^{15}}{3378375}-\frac{970432 u^{13}}{30405375}-\frac{1574008 u^{11}}{1216215}-\frac{5888416 u^9}{675675}-\frac{78549157 u^7}{10135125}+\frac{202117
   u^5}{1081080}-\frac{10027 u^3}{360360}\right)$\\
&$+\nu ^4 \left(\frac{64 u^{15}}{289575}+\frac{221696 u^{13}}{2027025}+\frac{24551302 u^{11}}{6081075}+\frac{30098609
   u^9}{1351350}+\frac{24703429 u^7}{1474200}+\frac{667423 u^5}{675675}-\frac{80539 u^3}{360360}\right)$\\
&$+\nu ^3 \left(-\frac{1024 u^{15}}{4343625}-\frac{559648
   u^{13}}{4343625}-\frac{14693678 u^{11}}{3378375}-\frac{306099041 u^9}{20270250}+\frac{199552117 u^7}{40540500}+\frac{5344433 u^5}{5405400}-\frac{13891 u^3}{32760}\right)$\\
&$+\nu ^2
   \left(\frac{128 u^{15}}{1126125}+\frac{2033672 u^{13}}{30405375}+\frac{126720793 u^{11}}{60810750}-\frac{22992913 u^9}{9009000}-\frac{475563619 u^7}{11583000}-\frac{5059069
   u^5}{1801800}-\frac{271 u^3}{12870}\right)$\\
&$+\nu  \left(-\frac{256 u^{15}}{10135125}-\frac{160096 u^{13}}{10135125}-\frac{4015859 u^{11}}{8687250}+\frac{350867149
   u^9}{81081000}+\frac{198258107 u^7}{6237000}+\frac{12539239 u^5}{900900}-\frac{1389263 u^3}{72072}\right)$\\
&$+\frac{64 u^{15}}{30405375}+\frac{14032 u^{13}}{10135125}+\frac{330643
   u^{11}}{8687250}-\frac{66302989 u^9}{81081000}-\frac{1304921 u^7}{216216}-\frac{1609339 u^5}{180180}+\frac{783883 u^3}{360360}  $\\
$A_{01}^{(9)}$ &$ \nu ^5 \left(\frac{137728 u^{14}}{30405375}+\frac{1630144 u^{12}}{3378375}+\frac{20156672 u^{10}}{2764125}+\frac{14561504 u^8}{921375}+\frac{16872197 u^6}{10135125}+\frac{20077
   u^4}{98280}\right)$\\
&$+\nu ^4 \left(-\frac{13184 u^{14}}{868725}-\frac{9757312 u^{12}}{6081075}-\frac{124302352 u^{10}}{6081075}-\frac{28373797 u^8}{810810}-\frac{2263879
   u^6}{737100}-\frac{159317 u^4}{5405400}\right)$\\
&$+\nu ^3 \left(\frac{75008 u^{14}}{4343625}+\frac{8051936 u^{12}}{4343625}+\frac{26744852 u^{10}}{1447875}+\frac{43044643
   u^8}{6756750}-\frac{63587191 u^6}{20270250}-\frac{14124959 u^4}{5405400}\right)$\\
&$+\nu ^2 \left(-\frac{265472 u^{14}}{30405375}-\frac{751736 u^{12}}{779625}-\frac{13442798
   u^{10}}{2338875}+\frac{519548429 u^8}{11583000}+\frac{1305091549 u^6}{40540500}-\frac{62753 u^4}{50050}\right)$\\
&$+\nu  \left(\frac{61184 u^{14}}{30405375}+\frac{6955744
   u^{12}}{30405375}+\frac{6686014 u^{10}}{30405375}-\frac{906835357 u^8}{27027000}-\frac{1624866443 u^6}{40540500}+\frac{6196783 u^4}{257400}\right)$\\
&$-\frac{5248
   u^{14}}{30405375}-\frac{87872 u^{12}}{4343625}+\frac{477758 u^{10}}{6081075}+\frac{155848333 u^8}{27027000}+\frac{22502021 u^6}{2702700}+\frac{3239563 u^4}{360360}  $\\
$A_{11}^{(9)}$ &$ \nu ^5 \left(-\frac{256 u^{15}}{3378375}-\frac{1984 u^{13}}{66825}-\frac{10816616 u^{11}}{10135125}-\frac{8043292 u^9}{1447875}-\frac{21410527 u^7}{10135125}+\frac{12005701
   u^5}{81081000}+\frac{7 u^3}{40}\right)$\\
&$+\nu ^4 \left(\frac{64 u^{15}}{289575}+\frac{619616 u^{13}}{6081075}+\frac{3994318 u^{11}}{1216215}+\frac{54889553 u^9}{4054050}+\frac{84250319
   u^7}{16216200}+\frac{5005309 u^5}{32432400}-\frac{15 u^3}{56}\right)$\\
&$+\nu ^3 \left(-\frac{1024 u^{15}}{4343625}-\frac{522656 u^{13}}{4343625}-\frac{105828574
   u^{11}}{30405375}-\frac{796579 u^9}{107250}+\frac{148683929 u^7}{40540500}+\frac{59417713 u^5}{40540500}-\frac{4973 u^3}{2520}\right)$\\
&$+\nu ^2 \left(\frac{128
   u^{15}}{1126125}+\frac{1902664 u^{13}}{30405375}+\frac{33080947 u^{11}}{20270250}-\frac{127094449 u^9}{27027000}-\frac{623545387 u^7}{27027000}+\frac{588428461
   u^5}{162162000}+\frac{3319 u^3}{693}\right)$\\
&$+\nu  \left(-\frac{256 u^{15}}{10135125}-\frac{90016 u^{13}}{6081075}-\frac{3084403 u^{11}}{8687250}+\frac{69433531
   u^9}{16216200}+\frac{1741716547 u^7}{81081000}+\frac{1558061443 u^5}{162162000}+\frac{1556747 u^3}{360360}\right)$\\
&$+\frac{64 u^{15}}{30405375}+\frac{13168 u^{13}}{10135125}+\frac{1737653
   u^{11}}{60810750}-\frac{6901903 u^9}{9009000}-\frac{13725913 u^7}{3003000}-\frac{50101837 u^5}{10810800}+\frac{155185 u^3}{72072} $\\
$A_{0{\rm exp}}^{(9)}$&$ \nu ^4 \left(\frac{128 u^{12}}{19305}-\frac{544 u^{11}}{3861}+\frac{36332 u^{10}}{27027}-\frac{953978 u^9}{135135}+\frac{61979 u^8}{2860}-\frac{2738621 u^7}{72072}+\frac{31847
   u^6}{936}-\frac{37731 u^5}{3640}-\frac{3368 u^4}{429}-\frac{160 u^3}{143}-\frac{160 u^2}{143}\right)$\\
&$+\nu ^3 \left(-\frac{256 u^{12}}{19305}+\frac{6568 u^{11}}{19305}-\frac{161146
   u^{10}}{45045}+\frac{87401 u^9}{4290}-\frac{7150567 u^8}{108108}+\frac{9193913 u^7}{77220}-\frac{1272923 u^6}{12012}+\frac{10317 u^5}{572}+\frac{1064914 u^4}{45045}-\frac{3656
   u^3}{1001}-\frac{3656 u^2}{1001}\right)$\\
&$+\nu ^2 \left(\frac{896 u^{12}}{96525}-\frac{76556 u^{11}}{289575}+\frac{44225903 u^{10}}{14189175}-\frac{2059303 u^9}{108108}+\frac{1488657707
   u^8}{22702680}-\frac{1986526567 u^7}{16216200}+\frac{12502472027 u^6}{113513400}-\frac{12421579 u^5}{793800}-\frac{226612271 u^4}{9459450}\right.$\\
&$\left.+\frac{2347742 u^3}{4729725}+\frac{17938
   u^2}{15015}\right)$\\
&$+\nu  \left(-\frac{256 u^{12}}{96525}+\frac{2666 u^{11}}{32175}-\frac{8243593 u^{10}}{7739550}+\frac{361685587 u^9}{48648600}-\frac{9193623619
   u^8}{340540200}+\frac{5908076453 u^7}{113513400}-\frac{5588921773 u^6}{113513400}+\frac{260971457 u^5}{113513400}-\frac{1930211 u^4}{1455300}\right.$\\
&$\left.-\frac{8469413 u^3}{4729725}+\frac{2531
   u^2}{429}\right)$\\
&$+\frac{128 u^{12}}{482625}-\frac{12959 u^{11}}{1447875}+\frac{320309987 u^{10}}{2554051500}-\frac{9740470933 u^9}{10216206000}+\frac{8793640367
   u^8}{2043241200}-\frac{497370187 u^7}{68108040}+\frac{2676313333 u^6}{340540200}-\frac{33055807 u^5}{48648600}+\frac{26267534 u^4}{14189175}$\\
&$-\frac{12487672 u^3}{14189175}-\frac{29984
   u^2}{15015} $\\
$A_{1{\rm exp}}^{(9)}$&$ \nu ^4 \left(\frac{128 u^{12}}{19305}-\frac{2656 u^{11}}{19305}+\frac{6388 u^{10}}{5005}-\frac{289984 u^9}{45045}+\frac{110459 u^8}{5940}-\frac{37799 u^7}{1287}+\frac{271039
   u^6}{12870}-\frac{26309 u^5}{12870}+\frac{587 u^4}{2310}-\frac{256 u^3}{429}\right)$\\
&$+\nu ^3 \left(-\frac{256 u^{12}}{19305}+\frac{1288 u^{11}}{3861}-\frac{461122
   u^{10}}{135135}+\frac{153211 u^9}{8190}-\frac{2809663 u^8}{49140}+\frac{49969981 u^7}{540540}-\frac{34973837 u^6}{540540}+\frac{85873 u^5}{15015}-\frac{64121 u^4}{9009}-\frac{21992
   u^3}{4095}\right)$\\
&$+\nu ^2 \left(\frac{896 u^{12}}{96525}-\frac{75212 u^{11}}{289575}+\frac{42399673 u^{10}}{14189175}-\frac{998111567 u^9}{56756700}+\frac{926299579
   u^8}{16216200}-\frac{1359512002 u^7}{14189175}+\frac{22460744 u^6}{363825}-\frac{41620637 u^5}{2579850}-\frac{261379 u^4}{2182950}+\frac{102218 u^3}{45045}\right)$\\
&$+\nu  \left(-\frac{256
   u^{12}}{96525}+\frac{1574 u^{11}}{19305}-\frac{87237077 u^{10}}{85135050}+\frac{262298081 u^9}{37837800}-\frac{8053606523 u^8}{340540200}+\frac{126854081 u^7}{3095820}-\frac{3665933
   u^6}{132300}+\frac{16528376 u^5}{14189175}-\frac{44665849 u^4}{5675670}-\frac{44887 u^3}{25025}\right)$\\
&$+\frac{128 u^{12}}{482625}-\frac{12767 u^{11}}{1447875}+\frac{61826833
   u^{10}}{510810300}-\frac{338263867 u^9}{378378000}+\frac{39545475419 u^8}{10216206000}-\frac{11167271519 u^7}{2043241200}+\frac{355268093 u^6}{56756700}+\frac{36303398
   u^5}{42567525}+\frac{392738209 u^4}{85135050}$\\
&$+\frac{115048 u^3}{75075}+\frac{57664 u^2}{15015}+\frac{57664 u}{15015} $\\
\hline
\end{tabular}
\end{ruledtabular}
\end{table*}

\begin{table*}
\caption{\label{Aabk}  Energy spectrum at 1-loop: list of coefficients at $p_\infty^{10}$.}
\begin{ruledtabular}
\begin{tabular}{ll}
$A_{00}^{(10)}$ &$ \nu ^4 \left(\frac{256 u^{13}}{57915}+\frac{6976 u^{11}}{9009}+\frac{75784 u^9}{7371}+\frac{1845434 u^7}{135135}-\frac{52597 u^5}{90090}+\frac{439 u^3}{6006}\right)$\\ 
&$+\nu ^3 \left(-\frac{512
   u^{13}}{57915}-\frac{5872 u^{11}}{2835}-\frac{1762652 u^9}{57915}-\frac{290096 u^7}{6237}-\frac{71849 u^5}{45045}+\frac{14939 u^3}{45045}\right)$\\
&$+\nu ^2 \left(\frac{1792
   u^{13}}{289575}+\frac{707648 u^{11}}{405405}+\frac{12619616 u^9}{405405}+\frac{4500796 u^7}{75075}+\frac{21961 u^5}{54054}-\frac{8227 u^3}{18018}\right)$\\
&$+\nu  \left(-\frac{512
   u^{13}}{289575}-\frac{76396 u^{11}}{135135}-\frac{8250169 u^9}{675675}-\frac{2987398 u^7}{96525}-\frac{6985439 u^5}{1351350}+\frac{618577 u^3}{90090}\right)$\\
&$+\frac{256
   u^{13}}{1447875}+\frac{210464 u^{11}}{3378375}+\frac{5428288 u^9}{3378375}+\frac{18193408 u^7}{3378375}+\frac{213376 u^5}{32175}-\frac{238592 u^3}{45045} $\\
$A_{01}^{(10)}$ &$ \nu ^4 \left(-\frac{256 u^{12}}{1485}-\frac{2720512 u^{10}}{405405}-\frac{10061096 u^8}{405405}-\frac{535952 u^6}{135135}-\frac{63839 u^4}{90090}\right)$\\
&$+\nu ^3 \left(\frac{24448
   u^{12}}{57915}+\frac{515360 u^{10}}{27027}+\frac{2095700 u^8}{27027}+\frac{5728292 u^6}{405405}+\frac{1889 u^4}{1287}\right)$\\
&$+\nu ^2 \left(-\frac{93056 u^{12}}{289575}-\frac{35855104
   u^{10}}{2027025}-\frac{189625064 u^8}{2027025}-\frac{21988364 u^6}{675675}+\frac{125039 u^4}{54054}\right)$\\
&$+\nu  \left(\frac{28192 u^{12}}{289575}+\frac{4312936
   u^{10}}{675675}+\frac{5980787 u^8}{135135}+\frac{511076 u^6}{19305}-\frac{35163433 u^4}{1351350}\right)$\\
&$-\frac{2944 u^{12}}{289575}-\frac{853312 u^{10}}{1126125}-\frac{20897704
   u^8}{3378375}-\frac{707008 u^6}{307125}+\frac{210176 u^4}{25025}   $\\
$A_{11}^{(10)}$ &$  \nu ^4 \left(\frac{256 u^{13}}{57915}+\frac{279872 u^{11}}{405405}+\frac{587992 u^9}{81081}+\frac{250234 u^7}{57915}-\frac{27 u^5}{70}-\frac{u^3}{2}\right)$\\
&$+\nu ^3 \left(-\frac{512
   u^{13}}{57915}-\frac{13744 u^{11}}{7371}-\frac{267628 u^9}{12285}-\frac{7207736 u^7}{405405}-\frac{14818 u^5}{57915}+\frac{79 u^3}{35}\right)$\\
&$+\nu ^2 \left(\frac{1792
   u^{13}}{289575}+\frac{3218816 u^{11}}{2027025}+\frac{6672112 u^9}{289575}+\frac{56632012 u^7}{2027025}-\frac{7840333 u^5}{1351350}-\frac{215 u^3}{126}\right)$\\
&$+\nu  \left(-\frac{512
   u^{13}}{289575}-\frac{209812 u^{11}}{405405}-\frac{298139 u^9}{32175}-\frac{348818 u^7}{19305}-\frac{2732953 u^5}{270270}-\frac{5609 u^3}{770}\right)$\\
&$+\frac{256
   u^{13}}{1447875}+\frac{580768 u^{11}}{10135125}+\frac{65248 u^9}{51975}+\frac{2665328 u^7}{675675}+\frac{5100736 u^5}{482625}+\frac{260096 u^3}{45045}$\\
$A_{0{\rm exp}}^{(10)}$&$0  $\\
$A_{1{\rm exp}}^{(10)}$&$ 0 $\\
\hline
\end{tabular}
\end{ruledtabular}
\end{table*}

\section{Concluding remarks}
In this paper, we have derived the energy emitted per unit frequency and per unit solid angle, using recent results for the gravitational waveform associated with gravitational waves emitted during a two-body scattering process at 1-loop order, i.e. $h \sim G^3$, and at 7.5PN accuracy. We have then converted this information into two related quantities: the non-linear memory and the radiated-energy spectrum.

This computation shows the importance of analytic results for gravitational waveforms. 
Any improvement on the state-of-the-art results provides a path to the extraction of several simpler quantities which describe as gravitational systems radiate. This information complements other state-of-the-art results focusing on the extraction of global quantities, i.e., quantities integrated over the frequencies and the celestial sphere), such as the total emitted energy and the scattering angle. In general, obtaining waveform-related analytic results at high PN orders is challenging. Nevertheless, whenever feasible, it provides valuable insight into effects, such as high-velocity order contributions to tails and radiation-reaction. For example, the spectrum computed in this work, has an undoubted importance in the study of the analytic continuation between unbound and bound quantities, as it is closely related to the extraction of the local action \cite{Dlapa:2024cje,Bini:2024tft}. Moreover, we think that our results will also  serve as valuable benchmarks for future PN computations.

\section*{Acknowledgments}

The authors thank Julio Parra Martinez for informative discussions.
D.B. acknowledges membership to the Italian Gruppo Nazionale per
la Fisica Matematica (GNFM) of the Istituto Nazionale
di Alta Matematica (INDAM), as well as the hospitality
and the highly stimulating environment of the Institut
des Hautes Etudes Scient\'if\'iques.
A.G. is grateful to the Istituto
per le Applicazioni del Calcolo \lq\lq M. Picone," CNR, Rome (IT) for
past support and hospitality during the development of
the present project.
This research was partially supported by the
2021 Balzan Prize for Gravitation: Physical and Astrophysical Aspects, awarded to T. Damour.

\end{document}